\documentclass[longauth]{aa}  

\usepackage{graphicx}
\usepackage{gensymb}

\usepackage{txfonts}
\usepackage[flushleft]{threeparttable}
\renewcommand{\arraystretch}{1.1}
\usepackage[hidelinks,colorlinks=true,linkcolor=blue,citecolor=blue]{hyperref}
\usepackage{xcolor}
\usepackage{longtable}
\usepackage{lipsum}
\usepackage{adjustbox}
\usepackage{stfloats}
\usepackage{scrextend}
\usepackage{amssymb}
\usepackage{amsmath}
\usepackage{mathrsfs}
\usepackage{multicol}
\usepackage{booktabs}
\usepackage{moresize}
\usepackage{orcidlink}
\usepackage{xspace}
\usepackage{placeins}

\newcommand{\feh}{\mbox{[Fe/H]}\xspace}

\newcommand{\mh}{\mbox{[M/H]}\xspace}

\newcommand{\teff}{\ensuremath{T_{\rm eff}}}

\newcommand{\gccc}{\mbox{g\,cm$^{-3}$}\xspace}

\newcommand{\me}{\mbox{$\it{M_{\rm \mathrm{\oplus}}}$}\xspace}
\newcommand{\re}{\mbox{$\it{R_{\rm \mathrm{\oplus}}}$}\xspace}

\newcommand{\msol}{\mbox{$\it{M_\mathrm{\odot}}$}\xspace}
\newcommand{\rsol}{\mbox{$\it{R_\mathrm{\odot}}$}\xspace}
\newcommand{\se}{\mbox{$\it{S_{\rm \mathrm{\oplus}}}$}\xspace}

\newcommand{\lsol}{\mbox{$\it{L_\odot}$}\xspace}

\begin{document}

   \title{Densities of small planets around the M dwarfs TOI-4336~A and TOI-4342 with ESPRESSO}
    \subtitle{Three sub-Neptunes, one super-Earth, and a Neptune-mass candidate}
\titlerunning{TOI-4336~A and TOI-4342 systems}
\authorrunning{L. Parc, et al.}
   \author{L\'ena Parc\inst{1,*}, 
    Charles Cadieux\inst{1,2},
    Nolan Grieves\inst{1}, 
    Fran\c{c}ois Bouchy\inst{1}, 
    Alexandrine L'Heureux\inst{2},
    Caroline Dorn\inst{3},
    Marie-Luise Steinmeyer\inst{3},
    Elisa Delgado-Mena\inst{4,5},
    René Doyon\inst{2},
    Yolanda G.C. Frensch\inst{1},
    Romain Allart\inst{2},
    Etienne Artigau\inst{2},
    Nicola Astudillo-Defru\inst{6},
    Xavier Bonfils\inst{7},
    Yann Carteret\inst{1},
    Ryan Cloutier\inst{8},
    Marion Cointepas\inst{1,7},
    Karen A.\ Collins\inst{9},
    Jose Renan De Medeiros\inst{10},
    Xavier Delfosse\inst{7},
    Xavier Dumusque\inst{1},
    Tianjun~Gan\inst{11},
    Jonay I. Gonz\'alez Hern\'andez\inst{12,13}, 
    Ravit Helled\inst{14},
    Monika Lendl\inst{1},
    Lucile Mignon\inst{7},
    Angelica Psaridi\inst{15,16},
    Nuno C. Santos\inst{5,17},
    Richard P. Schwarz\inst{9},
    Julia Venturini\inst{1}
          }

   \institute{
       \inst{1} Observatoire de Gen\`eve, D\'epartement d’Astronomie, Universit\'e de Gen\`eve, Chemin Pegasi 51, 1290 Versoix, Switzerland\\
       \inst{2} Institut Trottier de recherche sur les exoplan\`etes, D\'epartement de Physique, Universit\'e de Montr\'eal, Montr\'eal, Qu\'ebec, Canada \\
       \inst{3} Institute for Particle Physics and Astrophysics, ETH Zürich,  Otto-Stern-Weg 5, 8093 Zürich, Switzerland. \\
       \inst{4} Centro de Astrobiolog\'ia (CAB), CSIC-INTA, Camino Bajo del Castillo s/n, 28692, Villanueva de la Ca\~nada (Madrid), Spain.\\
       \inst{5} Instituto de Astrof\'isica e Ci\^encias do Espa\c{c}o, Universidade do Porto, CAUP, Rua das Estrelas, 4150-762 Porto, Portugal. \\
       \inst{6} Departamento de Matem\'atica y F\'isica Aplicadas, Universidad Cat\'olica de la Sant\'isima Concepci\'on, Alonso de Rivera 2850, Concepci\'on, Chile. \\
       \inst{7} Univ. Grenoble Alpes, CNRS, IPAG, F-38000 Grenoble, France. \\
       \inst{8} Department of Physics and Astronomy, McMaster University, 1280 Main St W, Hamilton, ON, L8S 4M1, Canada. \\
       \inst{9} Center for Astrophysics \textbar \ Harvard \& Smithsonian, 60 Garden Street, Cambridge, MA 02138, USA. \\
       \inst{10} Departamento de F\'isica Te\'orica e Experimental, Universidade Federal do Rio Grande do Norte, Campus Universit\'ario, Natal, RN, 59072-970, Brazil.\\
       \inst{11} Department of Astronomy, Westlake University, Hangzhou 310030, Zhejiang Province, China. \\
       \inst{12} Instituto de Astrof\'isica de Canarias (IAC), Calle V\'ia L\'actea s/n, 38205 La Laguna, Tenerife, Spain.\\
        \inst{13} Departamento de Astrof\'isica, Universidad de La Laguna (ULL), 38206 La Laguna, Tenerife, Spain.\\
       \inst{14} Center for Theoretical Astrophysics and Cosmology, Institute for Computational Science, University of Zürich, Winterthurerstrasse 190, 8057 Zürich, Switzerland. \\
       \inst{15} Institute of Space Sciences (ICE, CSIC), Carrer de Can Magrans S/N, Campus UAB, Cerdanyola del Valles, E-08193, Spain.\\
        \inst{16} Institut d’Estudis Espacials de Catalunya (IEEC), 08860 Castelldefels (Barcelona), Spain.\\
        \inst{17} Departamento de F\'isica e Astronomia, Faculdade de Ci\^encias, Universidade do Porto, Rua do Campo Alegre, 4169-007 Porto, Portugal. \\
        \inst{*}\email{lena.parc@unige.ch}    
             }

   \date{Received 28 November 2025; revised 30 January 2026; accepted 9 February 2026}

  \abstract{Characterizing the masses, radii, and compositions of small planets orbiting M dwarfs is key to understanding their formation and identifying the best targets for atmospheric follow-up with facilities such as JWST.}{We present the characterization of two planetary systems orbiting the M dwarfs TOI-4336~A (M3.5V) and TOI-4342 (M0V), each hosting two transiting planets previously validated with TESS and ground-based observations.}{We refined the photometry of the TOI-4342 system using TESS and LCOGT data, and characterized the host stars with NIRPS and ESPRESSO spectroscopy. High-precision ESPRESSO radial velocities (RVs) allowed us to constrain the planetary masses and investigate their potential compositions.}{The TOI-4336~A system is composed of a sub-Neptune with a period of 16.34 days, a radius of $2.14 \pm 0.08$~\re, and a mass of $3.33 \pm 0.36$~\me, along with an inner super-Earth on a 7.59-day orbit with a radius of $1.25 \pm 0.07$~\re and a mass of $1.55 \pm 0.13$~\me. The TOI-4342 system hosts two sub-Neptunes of similar sizes ($2.33 \pm 0.09$~\re and $2.35 \pm 0.09$~\re), with periods of 5.54 and 10.69 days. Their masses are measured to be $7.3 \pm 1.3$~\me and $4.8 \pm 1.4$~\me, respectively. The RVs also reveal a planet candidate around TOI-4342, most likely non-transiting, with a period of $47.5$ days and a minimum mass of $17.8 \pm 3.0$~\me.}{With precise radii and masses, we derived bulk densities and explored possible compositions. The TOI-4336~A sub-Neptune and super-Earth have densities of $1.87 \pm 0.30$ and $4.35 \pm 0.79$~g~cm$^{-3}$, while the two similar-sized sub-Neptunes in TOI-4342 show distinct densities of $3.18 \pm 0.67$ and $2.01 \pm 0.63$~g~cm$^{-3}$. Using an inference model, we find that TOI-4336~A~b, TOI-4342~b, and TOI-4342~c have an atmosphere mass fraction (AMF) of $\sim 3.7\%$, $\sim 1.8\%$, and $\sim 2.9\%$, respectively, while the super-Earth TOI-4336~A~c could contain $\sim 2\%$ of water or have a core-to-mass fraction (CMF) of $\sim 31\%$. All four planets are excellent targets for future atmospheric characterization with JWST, and their multi-planet nature makes them especially interesting for comparative planetology. Notably, TOI-4336~A~b stands out as one of the best-known targets in its size and temperature regime, with a transmission spectroscopy metric (TSM) of 138, comparable to benchmark planets such as K2-18~b and LHS~1140~b. Its inner sibling, TOI-4336~A~c, may also be of interest for emission spectroscopy and exploring the “cosmic shoreline”, similarly to the Rocky Worlds DDT JWST program.}


   \maketitle

\section{Introduction}

M dwarfs constitute the most numerous stellar population in the Milky Way \citep{Henry2006,Winters2015,Reyle2021}. Over the past decade, they have emerged as key targets for exoplanet studies, as multiple surveys have revealed that these low-mass stars host a high frequency of small planets \citep{Dressing2013,Bonfils2013,Dressing2015,Mignon2025}. Their small radii, low masses, and faint luminosities make them especially favorable for detecting and characterizing small and temperate planets through transits and radial velocities (RVs), enabling access to parameter regimes that remain challenging for larger, hotter stars.

The Transiting Exoplanet Survey Satellite (TESS) mission \citep{Ricker2014} has greatly enlarged the census of transiting planets around nearby M dwarfs. Because TESS monitors bright, nearby stars, its discoveries are particularly well suited for detailed characterization by follow-up observations from the ground. In the PlanetS catalog\footnote{\url{https://dace.unige.ch/exoplanets/}} of transiting planets with robust and precise mass and radius measurements \citep[$\sigma_M/M < 25\%$; $\sigma_R/R <8\%$;][]{Otegi2020,Parc2024}, $\sim$70\% of the confirmed M-dwarf planets (65 out of 91) were initially discovered by TESS, testifying to the mission's impact for this spectral type.

Precise measurements of radius and mass are required to derive bulk densities of planets, which in turn provide crucial constraints on internal structure and composition \citep{Dorn2015,Plotnykov2024}. Understanding the diversity of the compositions of M-dwarf planets is therefore central to understanding how their distinct formation environments shape planetary architectures. These stars differ markedly from FGK-type hosts: they experience prolonged pre-main-sequence phases \citep{Baraffe1998,Baraffe2015}, possess lower-mass protoplanetary disks \citep{Pascucci2016}, and exhibit stronger and longer-lived magnetic activity at young ages \citep{Ribas2005}.

While observational evidence indicates that close-in low-mass planets are more common around M dwarfs than around solar-type stars, the extent to which stellar environment affects planetary composition remains uncertain. \citet{Cloutier2020} reported that the frequency of short-period rocky planets increases with decreasing stellar mass, with the ratio of rocky to non-rocky planets being 6–30 times greater for mid-M dwarfs than for mid-K dwarfs. Likewise, \citet{Ment2023} found a terrestrial-to–sub-Neptune ratio of 14:1 for late-M dwarfs compared to GK stars. Theoretical work by \citet{Kubyshkina2021} suggests that atmospheric escape is more efficient around low-mass stars at a fixed equilibrium temperature, potentially contributing to the prevalence of rocky planets in M-dwarf systems. 

Despite the evidence suggesting that M dwarfs tend to form more rocky planets, recent statistical studies \citep{Parc2024, Parc2025} present small sub-Neptunes (1.8~\re < $R_p$ < 2.9~\re) orbiting M dwarfs to have lower densities compared to their FGK counterparts. Such low densities may be accounted for if these planets are ice-rich and formed beyond the snow line before migrating inward \citep[e.g.,][]{Alibert2017,Venturini2020,Venturini2024,Burn2021,Burn2024}. However, the current sample of well-characterized sub-Neptunes around M dwarfs remains limited (22 compared to 84 around FGK dwarfs). Expanding this sample is essential to determine whether this apparent low-density trend persists and to better constrain the diversity of planetary compositions around M dwarfs.

Precise mass and radius are also required for atmospheric studies, since the atmospheric scale height depends on the planet’s surface gravity \citep{Batalha2019}. The James Webb Space Telescope (JWST) is beginning to demonstrate its ability to probe the atmospheric composition of super-Earths and sub-Neptunes, confirming their broad compositional diversity \citep[e.g.,][]{Benneke2024,Piaulet2024,Alderson2024,Gressier2024,Schmidt2025,Malsky2025,Ahrer2025}.

In this context, we present the characterization of two planetary systems orbiting M dwarfs, TOI-4336~A and TOI-4342, contributing to the growing body of knowledge on the diversity and composition of planets in these systems. TOI-4336~A hosts a validated planet reported by \citet{Timmermans2024}, a 2.12~\re\ sub-Neptune on a 16.34-day orbit around an M3.5V star. The star is part of a triple system, with an equal-mass secondary at a projected separation of $\sim$140~au, which is an interesting configuration for formation studies. Receiving only 1.5 times the Earth’s insolation, this planet has an equilibrium temperature of 308~K. 

\citet{Timmermans2024} also reported an additional planet candidate in the system, which has since been validated through a photometric follow-up campaign \citep{Timmermans2026}. On the other hand, \citet{Tey2023} validated two sub-Neptunes of similar sizes (2.27~\re\ and 2.42~\re) orbiting TOI-4342 with periods near a 2:1 ratio (5.54 and 10.69 days). 

In this paper, we confirm the planetary nature of these four planets by measuring their masses through radial velocities obtained with the ESPRESSO spectrograph. For TOI-4342, we also refine the radii of both transiting planets using additional TESS sectors and report an additional planet candidate detected in the radial velocities. The paper is organized as follows: in Sect.~\ref{sect:observations}, we present the space- and ground-based observations from TESS, LCO, ESPRESSO, and NIRPS. Sect.~\ref{sect:stellar_charac} describes the determination of host star parameters and the analysis of stellar activity using both ESPRESSO and NIRPS high-resolution spectra. In Sect.~\ref{sect:phot_rv_analysis}, we present the photometric and radial velocity analysis of both systems along with the results. Finally, in Sect.~\ref{sect:discussion}, we discuss the properties of the systems and summarize our conclusions in Sect.~\ref{sect:conclusions}.

\section{Observations}\label{sect:observations}
\subsection{TESS photometry}

TOI-4336~A and TOI-4342 were observed by TESS over 3 and 6 sectors, respectively, during its primary and extended missions. TOI-4336~A was observed in Sector 11 (13 April 2019 - 20 May 2019), Sector 38 (29 April 2021 - 26 May 2021), and Sector 64 (6 April 2023 - 4 May 2023). TOI-4342 was observed in Sector 13 (19 June 2019 - 17 July 2019), Sector 27 (5 July 2020 - 30 July 2020), Sector 66 (2 June 2023 - 1 July 2023), Sector 67 (1 July 2023 - 29 July 2023), Sector 93 (3 June 2025 - 29 June 2025) and Sector 94 (29 June 2025 - 25 July 2025).

Since TOI-4336~A has not been observed by TESS since the study of \citet{Timmermans2024}, we did not include it in our photometric analysis. In contrast, TOI-4342 has benefited from four additional TESS sectors since the study by \citet{Tey2023}.

For our analysis, we made use of both the Simple Aperture Photometry \citep[SAP;][]{Twicken2010,Morris2020} and the Presearch Data Conditioning SAP \citep[PDCSAP;][]{Stumpe2012,Stumpe2014,Smith2012} light curves, as provided by the TESS Science Processing Operations Center \citep[SPOC;][]{Jenkins2016} at NASA Ames Research Center. For both targets, we used the SAP data to analyze stellar activity, since it preserves the photometric variability associated with stellar rotation. For TOI-4342, we employed the PDCSAP data for the transit modeling, as it accounts for dilution from contaminating sources within the TESS aperture.

\subsection{Ground-based photometry}

Between May 30 and September 14, 2021 (UT), four transits of TOI 4342~b and five transits of TOI 4342~c were observed in the Sloan i' band using the 1.0-m telescopes of the Las Cumbres Observatory Global Telescope (LCOGT) network \citep{Brown2013}. As \citet{Tey2023}, we considered seven out of the nine observed transits, excluding two (of planet c) that were truncated due to poor weather conditions. Observations were conducted at the Siding Spring Observatory (SSO), the South African Astronomical Observatory (SAAO), and the Cerro Tololo Inter-American Observatory (CTIO) nodes of LCOGT, as summarized in Table~\ref{tab:GB_transits_TOI4342}. The 1-m telescopes are equipped with 4096 $\times$ 4096 SINISTRO cameras, providing an image scale of 0.389'' per pixel and a 26' $\times$ 26' field of view. Images were calibrated using the standard LCOGT \texttt{BANZAI} pipeline \citep{McCully2018} and photometric measurements extracted using \texttt{AstroImageJ} \citep{Collins2017}. 

\begin{table}[h]
\small
\caption{Ground-based observations of TOI-4342.}
\centering
\renewcommand{\arraystretch}{1.1}
\setlength{\tabcolsep}{13pt}
\begin{center}
\begin{tabular}{lccc}
\hline\hline
\textbf{Planet} &\textbf{Facility} & \textbf{Date} & \textbf{Label} \\
\hline
    b & LCO-SAAO & 2021-06-19 & B1 \\
    & LCO-CTIO & 2021-07-12 & B2 \\
    & LCO-SSO & 2021-08-14 & B3 \\
    & LCO-CTIO & 2021-08-31 & B4 \\
\hline
    c & LCO-CTIO a$^{(a)}$ & 2021-08-13 & C1 \\
    & LCO-CTIO b$^{(a)}$ & 2021-08-13 & C2 \\
    & LCO-CTIO & 2021-09-14 & C3 \\
\hline
\end{tabular}
\begin{tablenotes}
\item
\textbf{Notes:} $^{(a)}$ Transit observed simultaneously by two distinct telescopes at CTIO.
\end{tablenotes}
\label{tab:GB_transits_TOI4342}
\end{center}
\end{table}

\subsection{ESPRESSO spectroscopy}\label{subsect:ESPRESSO_data}

We obtained 51 and 58 spectra of TOI-4336~A and TOI-4342, respectively, using ESPRESSO \citep{Pepe2021} on the 8.2-meter Very Large Telescope (VLT) at Paranal Observatory, Chile. The observations were conducted between April 2022 and March 2023 as part of ESO programs 109.2391.001 and 110.24AD.001 (PI: Grieves), dedicated to the characterization of sub-Neptunes orbiting M dwarfs. Each observation had an integration time of 1800 s, a median resolving power of 140,000 (with 2$\times$1 binning), and covered the 380–788 nm wavelength range.

We used spectra reduced with the ESPRESSO pipeline, retrieved from the ESO Archive\footnote{\url{https://archive.eso.org/scienceportal/home}}. Radial velocities were extracted with both the cross-correlation method (CCF) (using a M3 mask for TOI-4336~A and a M0 mask for TOI-4342) and the line-by-line (LBL; version 0.65.003) method of \citet{Artigau2022}, available as an open-source package. A preliminary telluric correction was applied within the LBL framework by fitting a TAPAS atmospheric model \citep{Bertaux2014}. This approach, similar to that of \citet{Allart2022}, has been shown to enhance ESPRESSO RV precision, especially for M-type stars \citep{Cadieux2024a,Cadieux2025}. The LBL method also provides activity indicators, including the Full-Width at Half-Maximum (FWHM) and differential temperature of the star ($\Delta T$), which are explained in \citet{Artigau2022} and \citet{Artigau2024}. We also used the classical CCF computation of FWHM, BIS span and $\log R'_\mathrm{HK}$. 

To remove potential wrong or low signal-to-noise measurements, we performed post-processing for the RVs in two steps. First, we applied 5$\sigma$ clipping directly on the RV time series to identify and reject outliers. Second, we removed points with RV uncertainties above the 95th percentile of the uncertainty distribution. As a result, 5 measurements were discarded for TOI-4336~A and 3 for TOI-4342. All the time series are publicly available through the DACE platform\footnote{\url{https://dace.unige.ch/}}.

\begin{figure*}[t]
\centering
    \includegraphics[width=0.48\textwidth]{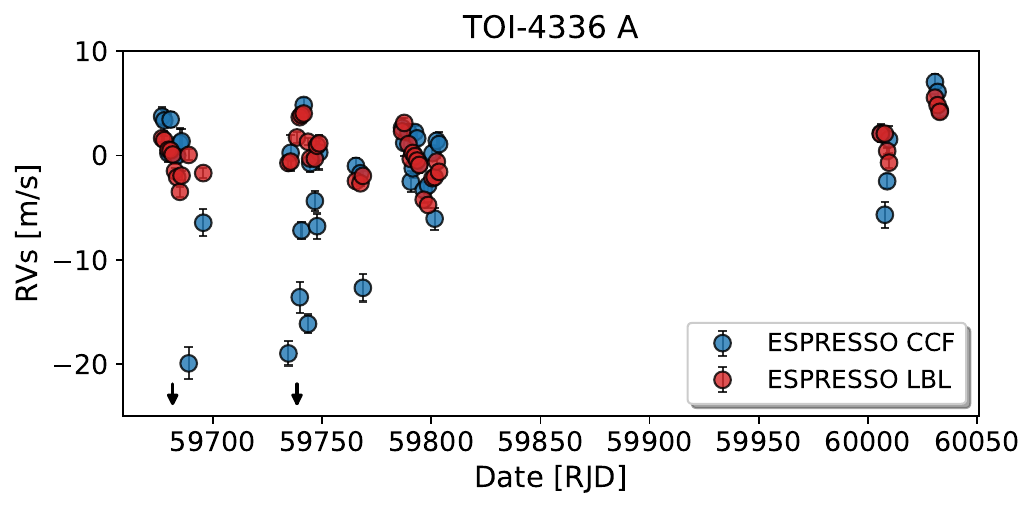}
   \includegraphics[width=0.48\textwidth]{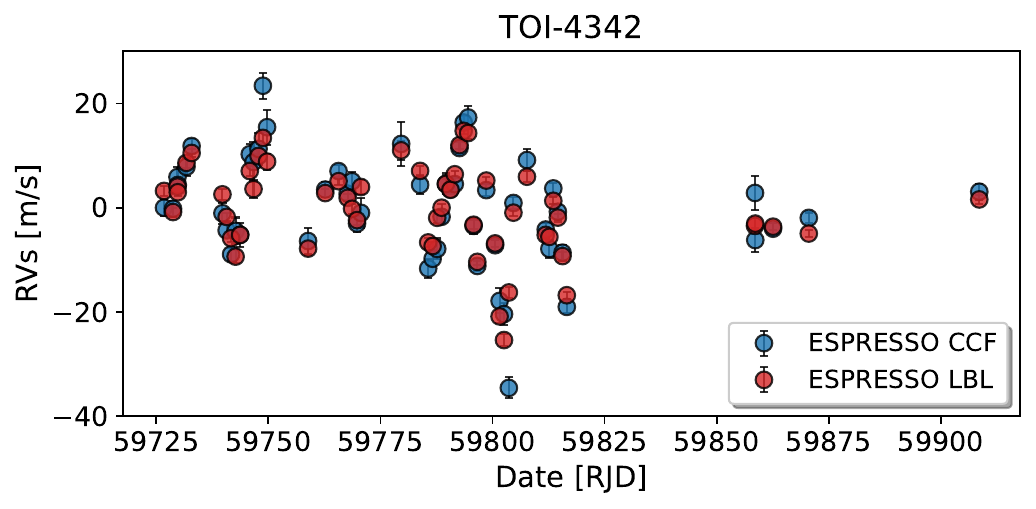}
     \caption{Radial velocity measurements of TOI-4336~A (left) and TOI-4342 (right) obtained with ESPRESSO, using the CCF method (blue points) and the LBL method (red points).}
     \label{fig:LBL_vs_CCF_RVs}
\end{figure*}

In Fig.~\ref{fig:LBL_vs_CCF_RVs}, we show the RVs of both targets derived using the CCF method (in blue) and the LBL method (in red). A significant improvement is obtained for both systems: for TOI-4336~A, the RV dispersion (robust standard deviation) decreases from 4.8 m s$^{-1}$ using the CCF to 2.2 m s$^{-1}$ with the LBL, while for TOI-4342 it decreases from 8.9 m s$^{-1}$ to 7.3 m s$^{-1}$. The LBL extraction also enables us to correct the outliers present in the TOI-4336~A dataset. In both cases, the RV photon-noise uncertainties are also reduced: from 0.89 m/s to 0.36 m/s for TOI-4336~A, and from 1.64 m/s to 0.89 m/s for TOI-4342. A similar gain over the CCF method was reported in the reanalysis of ESPRESSO data for the M4.5V dwarf LHS 1140 \citep{Cadieux2024a} and for the M3V dwarf L 98-59 \citep{Cadieux2025}. At the end, we used the radial velocities from the LBL method for our analysis in Sect.~\ref{sect:phot_rv_analysis}.

\subsection{NIRPS spectroscopy}

TOI-4336~A and TOI-4342 were observed from March 2024 to July 2024 and from August 2023 to October 2023, respectively, with the NIRPS \citep{Bouchy2025} echelle spectrograph at the ESO 3.6\,m telescope at La Silla Observatory in Chile. They have been observed as part of the NIRPS-GTO program, Follow-up of Transiting Planets subprogram (PID:111.254T.001, 112.25NS.001, 112.25NS.002; PI: Bouchy \& Doyon). They have been both observed with the High Efficiency (HE; R $\sim$ 75\,200, 0.9" fiber) mode of NIRPS with 2 exposures of 900 s per night of observations. We collected a total of 38 and 33 spectra, respectively for TOI-4336~A and TOI-4342, spread over 19 and 17 individual nights. For TOI-4336~A, one night (18-05-24) has been removed because the secondary component TOI-4336~B was pointed instead of the primary. The observations were processed with APERO \citep{cookAPEROPipelinEReduce2022}, the standard data reduction pipeline for the SPIRou near-infrared spectrograph \citep{donatiSPIRouNIRVelocimetry2020}, which is fully compatible with NIRPS. From the telluric-corrected APERO data products, we constructed a template spectrum of our targets, which we used to derive elemental abundances (Sect.~\ref{subsect:spectro_stellar}).

\section{Stellar characterization}\label{sect:stellar_charac}
\subsection{Spectroscopic parameters}\label{subsect:spectro_stellar}

We derived the spectroscopic stellar parameters of our targets by applying different techniques to the ESPRESSO and NIRPS spectra.
To obtain a high signal-to-noise spectrum, we first combined the ESPRESSO spectra using the IRAF task {\tt scombine}\footnote{IRAF is distributed by the National Optical Astronomy Observatories, operated by the Association of Universities for Research in Astronomy, Inc., under contract with the National Science Foundation, USA.}. The stellar atmospheric parameters were then estimated with the machine-learning code {\tt ODUSSEAS}\footnote{\url{https://github.com/AlexandrosAntoniadis/ODUSSEAS}} \citep{Antoniadis20,Antoniadis24}, which computes pseudo equivalent widths (EWs) for roughly 4000 optical spectral lines. These measurements are compared to a training set of 47 HARPS-observed M dwarfs, for which \feh\ values were determined from photometric relations \citep{Neves12} and \teff\ from interferometric measurements \citep{Khata21}. 

Secondly, the stellar abundances of chemical species were determined from the near-infrared NIRPS observations, following the methodology of \citet{jahandarComprehensiveHighresolutionChemical2024, jahandarChemicalFingerprintsDwarfs2025}. For each star, we combined the individual telluric-corrected spectra from the \texttt{APERO} reduction pipeline \citep[v0.7.292;][]{cookAPEROPipelinEReduce2022} into a single high-resolution spectrum. We then performed a series of $\chi^2$ fits on individual spectral lines using high-resolution stellar models from the PHOENIX ACES library \citep{Husser2013}, convolved to the spectral resolution of NIRPS. The method, originally developed for SPIRou data \citep{donatiSPIRouNIRVelocimetry2020}, benefits from $K$-band coverage to break the temperature–metallicity degeneracy. For NIRPS, which covers only the $YJH$ bands, this degeneracy can lead to a slight overestimation of $T_\mathrm{eff}$ for cooler stars ($<3500$\,K). To ensure consistency, we therefore adopted the $T_\mathrm{eff}$ values derived from the ESPRESSO analysis as fixed inputs for the NIRPS abundance determinations.

To account for the effect of the $T_\mathrm{eff}$ uncertainty, the abundance analysis was repeated at the two extremes of its $1\sigma$ confidence interval, using the same set of spectral lines as in the nominal analysis. For each chemical species, the difference between the results from these “extreme” cases and the nominal fit was added in quadrature to the line-to-line dispersion to obtain the final uncertainty estimates. The resulting elemental abundances and uncertainties are reported in Table~\ref{table:abundances_TOI4336TOI4342}.

\subsubsection{TOI-4336~A}

From the ESPRESSO spectra analysis, we obtained \teff\,$=3307\pm94$\,K and \feh\,$=-0.17\pm0.11$\,dex for TOI-4336~A. This is consistent with the values found in \citet{Timmermans2024}.

The combined NIRPS spectrum of TOI-4336\,A has a median S/N of 426. We adopted $\log g=5.0$, consistent with the measured $\log g=4.92\pm0.04$\,dex from \citet{Timmermans2024}. We also note that the method is calibrated to work within a range of 0.2\,dex. Using a PHOENIX model grid fixed at $T_\mathrm{eff}=3300$\,K, we find $[\mathrm{Fe/H}]=0.14\pm0.13$\,dex. This value is in mild ($1.8\sigma$) tension with the ESPRESSO metallicity. The Fe/Mg ratio of $0.62\pm0.53$ is consistent with the solar value ($0.79\pm0.10$; \citealt{Asplund2009}).

\subsubsection{TOI-4342}

From the ESPRESSO spectra analysis, we obtained \teff\,$=3866\pm92$\,K and \feh\,$=0.04\pm0.11$\,dex for TOI-4342. This is consistent with the \teff\ found in \citet{Tey2023}.

The combined NIRPS spectrum of TOI-4342 has a median S/N of 295. We adopted $\log g=4.75$, close to the measured $\log g=4.6878^{+0.0086}_{-0.0096}$ from \citet{Tey2023}. The analysis used a PHOENIX model grid fixed at $T_\mathrm{eff}=3860$\,K. The derived $[\mathrm{Fe/H}]$ is consistent with the ESPRESSO measurement, and the Fe/Mg ratio of $0.66\pm0.35$ agrees with the solar value within uncertainties.

\subsection{Stellar mass and radius}\label{subsect:stellar_mass_radius}

We derived the stellar masses and radii of TOI-4336~A and TOI-4342 using the empirically calibrated mass–luminosity and radius–luminosity relations for M dwarfs from \citet{Mann2015} and \citet{Mann2019}, respectively. This was done by combining the Gaia parallax with the 2MASS $K_s$-band magnitude to compute the distance and the absolute $K_s$-band magnitude ($M_K$). A Monte Carlo approach was used to propagate the observational uncertainties and to incorporate the intrinsic errors or the empirical relations, which are 2.2\% for the mass and 2.89\% for the radius. We then compared our derived stellar parameters with those reported in \citet{Timmermans2024,Tey2023}

\subsubsection{TOI-4336~A}

We found a stellar radius of $R_\star = 0.326 \pm 0.010~\rsol$ and a mass of $M_\star = 0.306 \pm 0.008~\msol$ for TOI-4336~A. While our radius measurement is consistent with that of \citet{Timmermans2024} ($0.330\pm0.015~\rsol$), our mass estimate differs significantly from their value ($0.331\pm0.010~\msol$). This discrepancy can be explained by them using the relation from \citet{Mann2015} despite the fact that \citet{Mann2019} is mentioned in the paper. We decided to use our stellar parameters estimates in this study as the radius and mass are consistent together with the effective temperature to a M3.5V dwarf according to the updated table 5 from \citet{Pecaut2013}.

\subsubsection{TOI-4342}

We found a stellar radius of $R_\star = 0.598 \pm 0.018~\rsol$ and a mass of $M_\star = 0.587 \pm 0.013~\msol$ for TOI-4342. Similarly, our radius measurement is consistent with that of \citet{Tey2023} ($0.599\pm0.013~\rsol$), but our mass estimate differs from their value ($0.6296\pm0.0086~\msol$). This discrepancy can be explained by the fact that they used the relations from \citet{Benedict2016} and not from \citet{Mann2019}. The difference between the relations is explored in \citet{Mann2019}. We decided to use our stellar parameters estimates in this study as the radius and mass are consistent together with the effective temperature to a M0V dwarf according to the updated table 5 from \citet{Pecaut2013}.

\subsection{Rotation period and activity indexes}\label{subsect:stellar_rot_activity}

We investigated the stellar activity of our targets by computing generalized Lomb-Scargle (GLS) periodograms for the radial velocities, activity indices, and TESS SAP photometry. From the LBL analysis, we used the RVs, the FWHM, and the $\Delta T_{3500K}$. From the CCF analysis, we used the FWHM, the BIS span, and the $\log R'_\mathrm{HK}$. The resulting periodograms are shown in Fig.~\ref{fig:TOI4336_periodograms}, where the orbital periods of the planets are indicated as vertical colored dashed lines.

\subsubsection{TOI-4336~A}

TOI-4336~A exhibits clear signs of stellar activity, with a highly significant peak at $\sim$36.7~days in the LBL RVs, and another peak at roughly half this period ($\sim$18~days) in the LBL FWHM and $\Delta T_{3500K}$ indicators (see Fig.~\ref{fig:TOI4336_periodograms}). No rotation period was reported in the TESS data by \citet{Timmermans2024}. However, when computing periodogram of the SAP flux from the three available sectors, we detect some power around $\sim$36~days, despite this being longer than a single TESS sector. No significant peaks are found in the CCF-based indicators, which is consistent with the strong improvement of the LBL method for the RVs (Sect.~\ref{subsect:ESPRESSO_data}).  
Despite the stellar activity, a significant peak is detected at the period of planet~b (16.3~days; pink dashed line). The peak corresponding to planet~c (7.6~days; light-blue dashed line) does not reach significance.
In addition, we recomputed $\log R'_\mathrm{HK}$ using the S-index, which measures the Ca~\textsc{ii} H \& K emission, and the $B-V$ color, as the ESPRESSO DRS had adopted an incorrect $B-V$ value. Following the relations from \citet{SuarezMascareno2015,SuarezMascareno2016}, we find a median value of $-5.65$. Using the relation from \citet{SuarezMascareno2018}, we then estimate the stellar rotation period to be $97^{+156}_{-57}$~days. This is somewhat longer than our observed value of 36.7~days but not excluded; however, the derivation of the S-index and $\log R'_\mathrm{HK}$ is known to be less reliable for late-M dwarfs, making such discrepancies expected.

\subsubsection{TOI-4342}

\citet{Tey2023} reported a rotation period of about 14~days for TOI-4342 based on TESS light curves. We confirmed this value with the inclusion of new TESS sectors (bottom panel of Fig.~\ref{fig:TOI4336_periodograms}). The LBL RVs and $\Delta T_{3500K}$, as well as the CCF FWHM and $\log R'_\mathrm{HK}$, also show a significant peak at this period ($\sim$14.9~days). In contrast, the LBL FWHM does not display any signal at this period, and the CCF BIS shows a peak at a shorter period ($\sim$4~days).  
In this case, the RVs are clearly dominated by stellar activity, and no significant peaks are detected at the orbital periods of the transiting planets, indicated as green and yellow dashed lines in the periodograms.
In addition, we find a median value of $\log R'_\mathrm{HK} = -4.64$, which is consistent with the classification of \citet{Henry1996} for an active star. Using the relation from \citet{SuarezMascareno2018}, we estimate the stellar rotation period from $\log R'_\mathrm{HK}$ and obtain a value of $18^{+24}_{-10}$~days, consistent with the $\sim$14-day period derived from the photometric and spectroscopic indicators.

\begin{figure*}[t]
\centering
    \includegraphics[width=\textwidth]{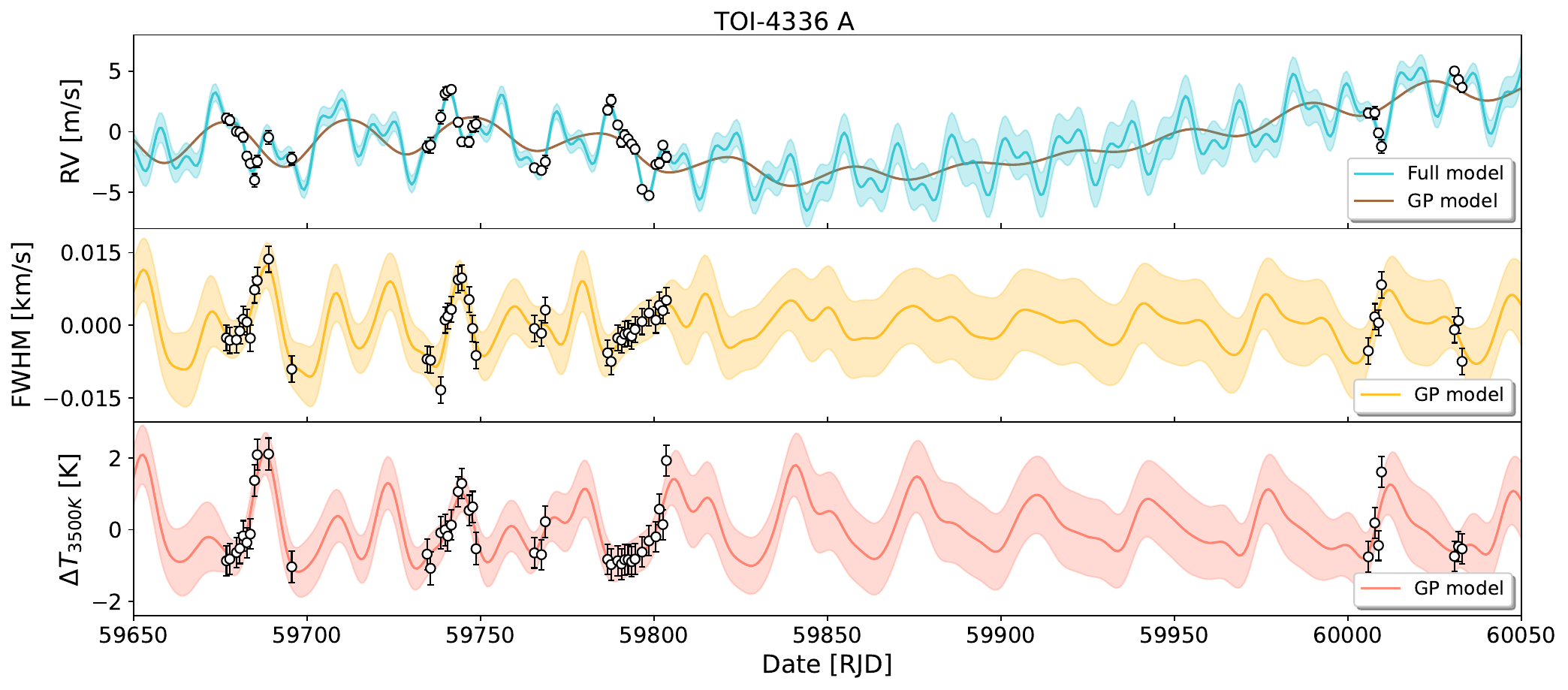}
     \caption{Time series of the RVs, FWHM, and $\Delta T_{3500\,\mathrm{K}}$ of TOI-4336~A after correction for the inferred offsets. The measurements are shown as white dots with error bars that include the fit jitter. In the top panel, the full RV + activity model with its 1$\sigma$ uncertainty is shown in blue, while the GP-only component is shown in brown. In the middle and bottom panels, the activity models are displayed in yellow and pink, respectively, with their associated shaded uncertainties.}
     \label{fig:RVs_full_model_4336}
\end{figure*}

\section{Photometric and radial velocity analysis}\label{sect:phot_rv_analysis}

We used \texttt{PyORBIT}\footnote{\url{https://github.com/LucaMalavolta/PyORBIT}} \citep{Malavolta2016,Malavolta2018} to model both the light curves and radial velocities of the two targets. This versatile Python framework enables the analysis of light curves, RVs, TTVs, as well as stellar activity using Gaussian Processes (GPs). We used the PyDE\footnote{\url{https://github.com/hpparvi/PyDE}} + emcee \citep{ForemanMackey2013} PyORBIT set-up. For each fit, we employed 4 × n$_\mathrm{dim}$ walkers, where n$_\mathrm{dim}$ is the dimensionality of the parameter space, ran the chains for a number of iterations tailored to the problem at hand, and applied a thinning factor of 100. Convergence was assessed by using the Gelman–Rubin statistic \citep{GelmanRubin1992}, adopting a threshold of $\hat{R} = 1.01$. In addition, we performed an autocorrelation analysis of the chains: if their length exceeded 50 times the estimated autocorrelation time, we deemed the chains converged. 

\subsection{TOI-4336~A}\label{subsect:RVs_TOI4336}

We fit a two-planet model to our ESPRESSO LBL RVs, adopting an uniform prior on the semi-amplitude, $K \sim \mathcal{U}(0,10)~\mathrm{m\,s^{-1}}$. For the orbital periods and times of inferior conjunction, we imposed Gaussian priors based on \citet{Timmermans2026}. Stellar activity was modeled using a Gaussian Processes (GP) framework \citep{Rajpaul2015}, in which we simultaneously fit the LBL RVs, LBL FWHM, and LBL $\Delta T_{3500K}$, as they are the activity indicators showing powers in their periodogram. Our GP model can be written as

\begin{multline*} 
RV = A_1 \, G_1(t) \\ 
\Delta T_{3500K} = A_2 \, G_2(t) \\ 
FWHM = B_2 \, G_2(t), \\ 
\end{multline*}

where $G_1(t)$ and $G_2(t)$ are two GPs with a quasi-periodic (QP) covariance kernel, as defined in \citet{Rajpaul2015},

\begin{equation} \label{eq:qp_kernel}
\gamma_{i, j}^{(G_k, G_k)}=\exp \left\{-\frac{\sin ^2\left[\pi\left(t_i-t_j\right) / P_{\mathrm{rot}}\right]}{2 O_{\mathrm{amp},k}^2}-\frac{\left(t_i-t_j\right)^2}{2 P_{\mathrm{dec}}^2}\right\} ,
\end{equation}

where $P_{\mathrm{rot}}$ is the stellar rotation period, $O_{\mathrm{amp}}$ is the coherence scale, and $P_{\mathrm{dec}}$ is the active-region decay timescale. In practice, we employed the S+LEAF implementation of the exponential-sine-squared kernel corresponding to this definition \citep{Delisle2020,Delisle2022}. The GP hyper-parameters $P_{\mathrm{rot}}$ and $P_{\mathrm{dec}}$ are shared between $G_1(t)$ and $G_2(t)$, while separate coherence scales $O_{\mathrm{amp},k}$ were fit for the RVs and for the “photometric” indicators ($\Delta T_{3500K}$ and FWHM). This distinction is physically motivated: while RVs are primarily modulated at the stellar rotation period, the photometric indicators vary on half that period, as illustrated in Fig.~\ref{fig:TOI4336_periodograms}.

We adopted uniform priors of $\mathcal{U}(10,100)$ days for the stellar rotation period and $\mathcal{U}(40,300)$ days for the active-region decay timescale, along with log-uniform priors log$\mathcal{U}(0.01,10)$ for the two coherence scales \citep{Stock2023}. Uniform priors were also used for the amplitudes: $\mathcal{U}(0,10)$ m/s for the RVs, $\mathcal{U}(0,10)$ km/s for the FWHM, and $\mathcal{U}(0,10)$ K for $\Delta T_{3500\mathrm{K}}$. In addition, we fit a jitter and an offset term ($\sigma$, $\mu$) for each dataset.

We tested scenarios with eccentric or circular orbits for one or both planets by adopting the parametrization of \citet{Eastman13}, in which the fit parameters are ($\sqrt{e}\,\sin\omega$, $\sqrt{e}\,\cos\omega$), from which the eccentricity $e$ and argument of periastron $\omega$ are derived. We also explored the possibility of an additional third planet in the system.

We compared the Bayesian Information Criterion (BIC) values of the different models to identify the most significant configuration. The two-planet circular model provides the lowest BIC ($-73.95$), compared to $-54.69$ for a two-planet eccentric model, $-64.77$ when allowing eccentricity only for planet b, $-64.04$ when allowing eccentricity only for planet c, and $-59.32$ for a three-planet model. In all cases, the semi-amplitudes of planets b and c remain fully consistent across models. The three-planet model returns a third signal with a semi-amplitude consistent with zero and fails to converge on a well-defined orbital period. Still, the GP component in the RV model appears to capture a long-term variation that is not present in the activity indicators, potentially hinting at the presence of a long-period companion that cannot be recovered within our observational timespan. For the eccentric models, the inferred eccentricities remain statistically consistent with zero, with $e_b = 0.082^{+0.078}_{-0.057}$ and $e_c = 0.063^{+0.081}_{-0.045}$, in agreement with the photometric analysis from \citet{Timmermans2026}. We therefore adopt the two-planet circular model as the preferred solution. From our analysis, we derive $3\sigma$ upper limits on the eccentricities of planet b and planet c of 0.45 and 0.35, respectively.

\begin{figure}[t]
\centering
    \includegraphics[width=0.48\textwidth]{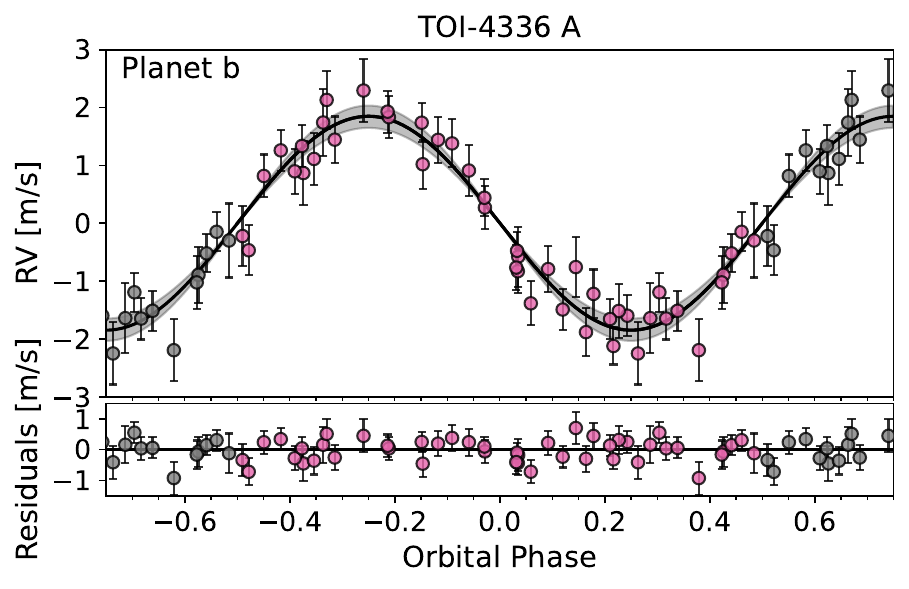}
    \includegraphics[width=0.48\textwidth]{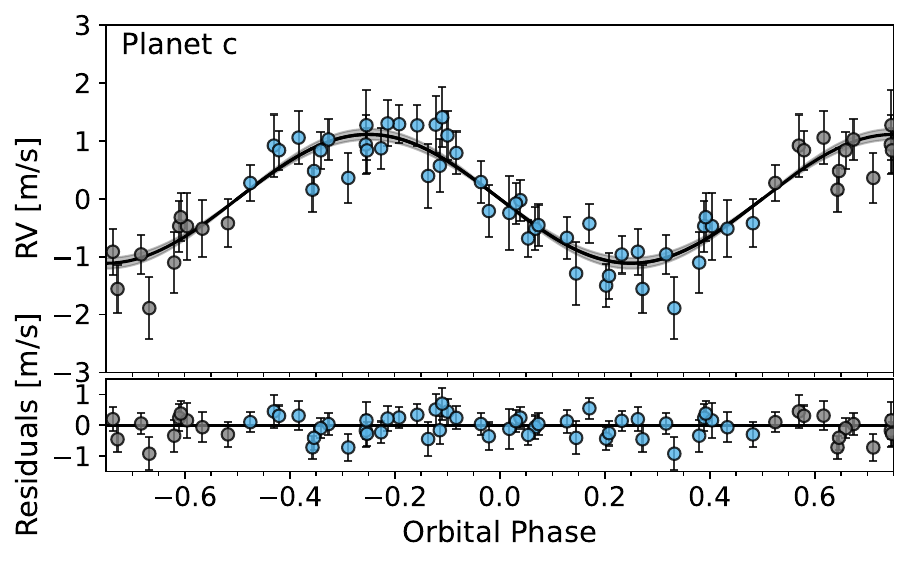}
     \caption{Phase-folded ESPRESSO RVs with the best-fit model from our analysis (Sect.~\ref{subsect:RVs_TOI4336}) and the corresponding residuals for TOI-4336~A~b (top) and TOI-4336~A~c (bottom). The errorbars include the fit jitter added in quadrature.}
     \label{fig:RVs_planets_folded_4336}
\end{figure}

The RV, FWHM, and $\Delta T_{3500K}$ time series, together with the best-fit model, are presented in Fig.~\ref{fig:RVs_full_model_4336}. The phase-folded RVs of the two planets are shown in Fig.~\ref{fig:RVs_planets_folded_4336}, while the priors and posteriors of the fit are listed in Table~\ref{tab:posteriors_RV_TOI4336}, and the derived planetary parameters are summarized in Table~\ref{tab:planetaryparams_TOI4336_TOI4342}.

Finally, we determined the masses of the two planets and re-derived their radius with the updated stellar parameters from Sect.~\ref{subsect:stellar_mass_radius}, thereby confirming their planetary nature. TOI-4336~A~b is a sub-Neptune with a radius of $2.14 \pm 0.08~R_\oplus$ and a mass of $3.33 \pm 0.36~M_\oplus$, orbiting its host star with a period of 16.34 days. With  TOI-4336~A~c is a super-Earth with a radius of $1.25 \pm 0.07~R_\oplus$ and a mass of $1.55 \pm 0.13~M_\oplus$, on an inner orbit with a period of 7.59 days. 

We also confirmed the stellar rotation period to be $35.5 \pm 1.1$ days from our GP modeling.

\begin{table*}[t]
\tiny
\caption{Fit and derived parameters for TOI-4336 A and TOI-4342 systems.}
\centering
\renewcommand{\arraystretch}{1.5}
\begin{center}
\begin{tabular}{lcc|ccc}
\hline\hline
 & \multicolumn{2}{c}{\textbf{TOI-4336 A}} & \multicolumn{3}{c}{\textbf{TOI-4342}} \\
\hline
\textbf{Parameter} &\textbf{Planet b} & \textbf{Planet c} &\textbf{Planet b} & \textbf{Planet c} & \textbf{Candidate d}\\
 \hline
    Orbital period, $\it{P_{\mathrm{orb}}}$ (days)\dotfill  & $^\dagger$16.336351 $\pm$ 0.000017 & $^\dagger$7.587266 $\pm$ 0.000012 & 5.5382592 $\pm$ 0.0000034 & 10.688662 $\pm$ 0.00015 & 47.5 $\pm$ 1.3 \\
    Time of conjunction, $T_{0}$ (RJD)\dotfill  & $^\dagger$59335.57275 $\pm$ 0.00046 & $^\dagger$59333.2931 $\pm$ 0.0015 & 58654.53479$^{+0.00084}_{-0.00093}$ & 58659.3486 $\pm$ 0.0017 & 58795 $\pm$ 27 \\
    Planet radius, $\it{R_{\mathrm{p}}}$ (\re)\dotfill  & 2.137 $\pm$ 0.080 & 1.251 $\pm$ 0.066 & 2.329$^{+0.086}_{-0.085}$  & 2.349$^{+0.094}_{-0.092}$ & - \\
    Planet mass, $\it{M_{\mathrm{p}}}$ (\me)\dotfill  & 3.33$^{+0.34}_{-0.37}$  &  1.55 $\pm$ 0.13 & 7.3 $\pm$ 1.3 & 4.8 $\pm$ 1.4 & $^\star$17.8$^{+2.9}_{-3.0}$ \\
    Planet density, $\rho_\mathrm{p}$ (g~cm$^{-3}$)\dotfill  & 1.87$^{+0.31}_{-0.28}$ & 4.35$^{+0.86}_{-0.71}$ &   3.18$^{+0.70}_{-0.63}$ &  2.01$^{+0.65}_{-0.60}$ & - \\
    RV semi-amplitude, $K$ ($\mathrm{m}\,\mathrm{s}^{-1}$) \dotfill & 1.85$^{+0.20}_{-0.18}$ & 1.114 $\pm$ 0.094 & 3.78$^{+0.65}_{-0.67}$ & 1.97$^{+0.57}_{-0.56}$ & 4.49$^{+0.73}_{-0.75}$\\
    Orbital inclination, $i$ ($^\circ$)\dotfill & $^\dagger$89.492 $\pm$ 0.093  &  $^\dagger$89.64 $\pm$ 0.26 &  88.83 $\pm$ 0.22 &  89.61 $\pm$ 0.26 & -  \\
    Scaled planetary radius, $R_{P}$/$R_{*}$   \dotfill & $^\dagger$0.0601 $\pm$ 0.0013 &  $^\dagger$0.0352 $\pm$ 0.0015 &  0.03571$^{+0.00073}_{-0.00075}$ &  0.03602$^{+0.00092}_{-0.00093}$ & -  \\
    Impact parameter, $\textit{b}$\dotfill  & $^\dagger$0.497$^{+0.071}_{-0.066}$ & $^\dagger$0.21$^{+0.12}_{-0.14}$ & 0.382$^{+0.062}_{-0.063}$ & 0.20$^{+0.12}_{-0.13}$ & -\\
    Semi-major axis, $\it{a}$ (au)\dotfill  & 0.08490$^{+0.00073}_{-0.00075}$ & 0.05092$\pm$0.00045 &0.0519$^{+0.00020}_{-0.00021}$ &0.0804$^{+0.00032}_{-0.00031}$ & 0.2154$^{+0.0040}_{-0.0043}$\\
    Eccentricity, $e$ \dotfill  & 0~($<$ 0.45, 3$\sigma$) & 0~($<$ 0.35, 3$\sigma$) & 0 & 0 & 0\\
    Argument of periastron, $\omega$ (deg) \dotfill  & 90 (fixed) &  90 (fixed) &  90 (fixed)  &  90 (fixed)  &  90 (fixed)  \\
    Transit duration, $T_{14}$ (hours) \dotfill  &  2.08 $\pm$ 0.11 & 1.74 $\pm$ 0.08 &  2.182 $\pm$ 0.024  &  2.854 $\pm$ 0.053  &  -  \\
    Insolation$^{(a)}$, $\it{S_{\mathrm{p}}}$ (\se)\dotfill  & 1.53 $\pm$ 0.14 & 4.24 $\pm$ 0.39 & 27.8$^{+3.6}_{-3.2}$ & 11.6$^{+1.5}_{-1.4}$ & 1.62 $\pm$ 0.16\\
    Equilibrium temperature$^{(a)}$, $T_{\mathrm{eq}}$ (K)\dotfill  &  309.4$^{+6.9}_{-7.4}$ & 399.5$^{+8.9}_{-9.5}$ & 639$^{+20}_{-19}$ & 514 $\pm$ 16 & 314$^{+7.6}_{-8.0}$ \\
    TSM $^{(b)}$\dotfill  & 138$^{+21}_{-16}$ & 12 $\pm$ 2  & 42$^{+11}_{-8}$  & 54$^{+23}_{-13}$  & -\\
    ESM $^{(b)}$\dotfill  & 0.82$^{+0.12}_{-0.11}$ & 1.01$^{+0.15}_{-0.13}$  & 3.86$^{+0.55}_{-0.49}$  & 1.95$^{+0.31}_{-0.27}$  & -\\
      \hline
\end{tabular}
\begin{tablenotes}
\item
\textbf{Notes:} $^{(a)}$ Insolation and equilibrium temperature are calculated as in \citet{Parc2024}, assuming global circulation and a Bond albedo of A$_\mathrm{B}$ = 0. $^{(b)}$ Transmission spectroscopy metric (TSM) calculated following \citet{Kempton2018}.$^\dagger$ Values from \citet{Timmermans2026}. $^\star$ Minimum mass.
\end{tablenotes}
\label{tab:planetaryparams_TOI4336_TOI4342}
\end{center}
\end{table*}

\subsection{TOI-4342}\label{subsect:analysis_TOI4342}

\subsubsection{Photometry analysis}\label{subsubsect:photometry_TOI4342}

We used \texttt{batman} \citep{Kreidberg2015_batman}, as implemented in \texttt{PyOrbit}, to fit the mid-transit times ($T_0$), orbital periods ($P$), planet-to-star radius ratios ($R_p/R_\star$), and impact parameters ($b$) of the two planets, along with the stellar density ($\rho_\star$), which was constrained by a Gaussian prior based on the results presented in Sect.~\ref{subsect:stellar_mass_radius}. We employed the TESS PDCSAP fluxes, which were further detrended for residual systematics and stellar activity using a Matérn 3/2 Gaussian Process kernel for each sector. To optimize computational efficiency, we binned the out-of-transit data to a one-hour cadence, while retaining the original TESS 120- or 20-second exposures for the in-transit windows ($T_0 \pm 10$ hours; see Fig.~\ref{fig:TESS_full_model_4342}). The ground-based photometry was detrended using airmass for all transits except B1, where we used the $y$-position of the centroid, and C3, where we used the sky/pixel flux. We derived the stellar limb-darkening (LD) coefficients and their uncertainties for each photometric filter using a quadratic LD law and the \texttt{LDCU}\footnote{\url{https://github.com/delinea/LDCU}} code \citep{Deline2022}. The \texttt{LDCU} code is a modified version of the Python routine by \citet{Espinoza2015}, which computes LD coefficients and their uncertainties from stellar intensity profiles while accounting for the uncertainties in the stellar parameters. These intensity profiles are generated using two libraries of synthetic stellar spectra: ATLAS \citep{Kurucz1979} and PHOENIX \citep{Husser_2013}. The LD coefficients obtained with \texttt{LDCU} were then adopted as Gaussian priors in our fit, assuming a slightly enlarged uncertainty of 0.1 to account for possible additional systematic errors. Finally, we added a jitter term to represent excess uncorrelated noise of each dataset. We adopted uniform priors for all other fit parameters.

\begin{figure*}[t]
  \centering
    \includegraphics[width=0.48\textwidth]{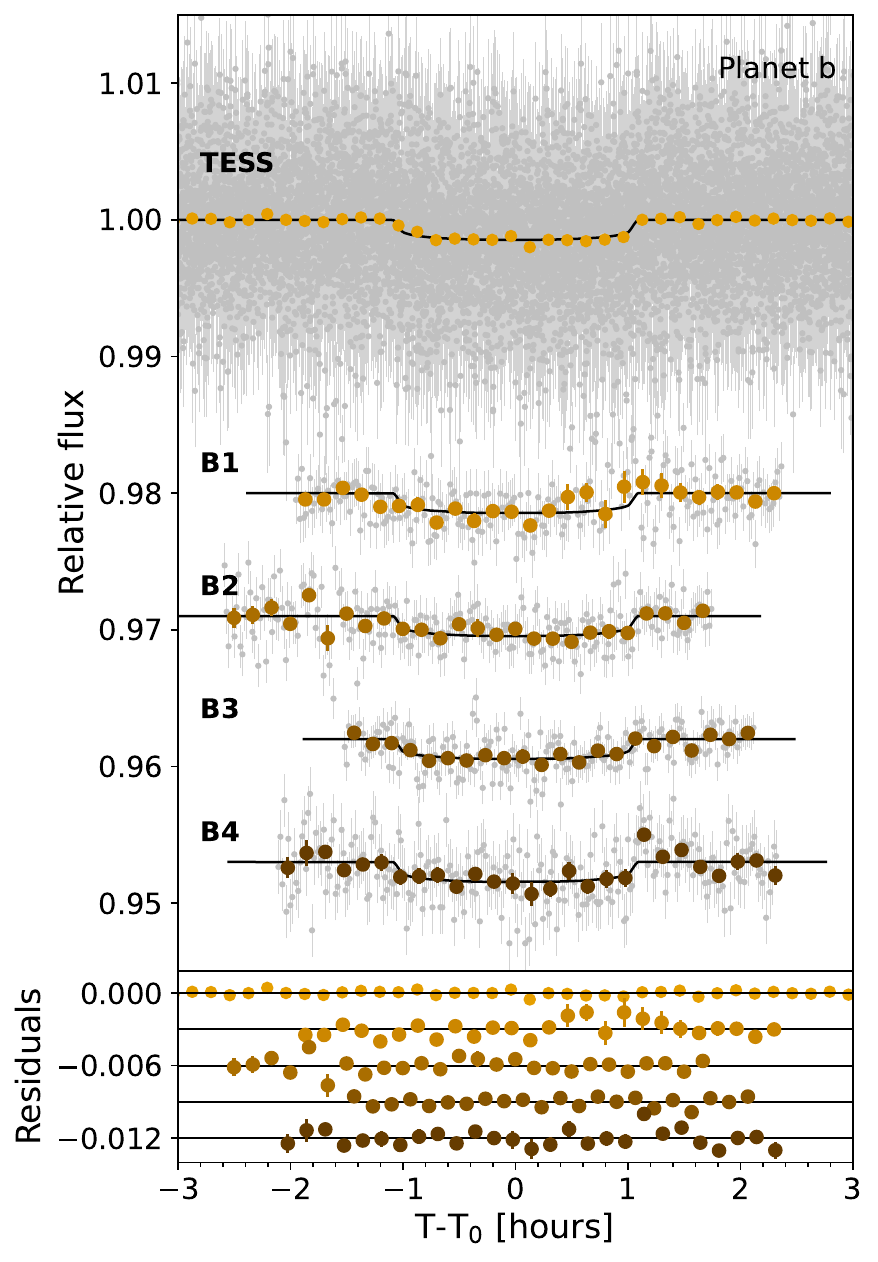}
    \includegraphics[width=0.48\textwidth]{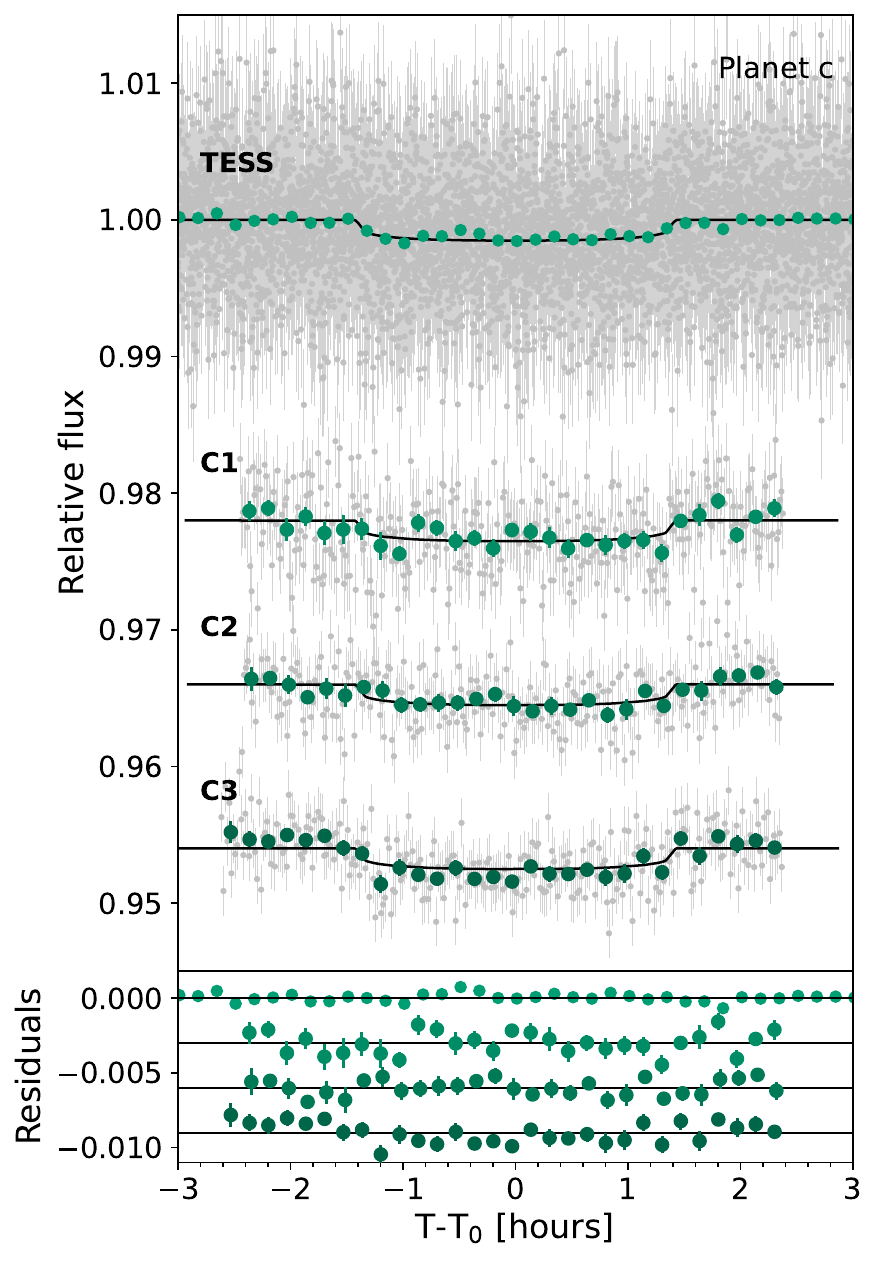}
  \caption{Phase-folded TESS and LCO light curves of TOI-4342 b (left) and TOI-4342 c (right), along with their residuals to the best-fit model from our analysis (Sect.~\ref{subsubsect:photometry_TOI4342}), shown as black lines. Colored circles represent the data binned in 10-minute intervals.} 
  \label{fig:TOI4342_LCs_phase_folded}
\end{figure*}

The full TESS light curves and best models of the six sectors are presented in Fig.~\ref{fig:TESS_full_model_4342}, and the phase-folded transits are plotted in Fig.~\ref{fig:TOI4342_LCs_phase_folded}. The priors and posteriors of the fit are listed in Table~\ref{tab:posteriors_RV_TOI4342}, and the derived planetary parameters are summarized in Table~\ref{tab:planetaryparams_TOI4336_TOI4342}.

Finally, we find that TOI-4342 b and c have very similar radii of 2.33 $\pm$ 0.09~\re and 2.35 $\pm$ 0.09~\re, respectively. 

Following \citet{Tey2023}, we also searched for evidence of transit-timing variations (TTVs) in the photometry of TOI-4342, given that the two transiting planets have a period ratio close to 2:1 ($P_c/P_b = 1.93$). We fit individual transit times from the TESS and LCO datasets using \texttt{PyOrbit} and compared them to the expected linear ephemeris in Fig.~\ref{fig:TTVs_4342}. No significant deviations were detected. The predicted TTV amplitude is only a few minutes, with a super-period of $\sim$150 days. However, the signal-to-noise ratio of individual TESS and ground-based transits is insufficient to reach this level of precision. The measured transit times show a scatter comparable to their uncertainties, with a few outliers that can be attributed to residual stellar activity affecting individual transits.

Ultimately, given the small expected amplitude, any undetected TTVs are unlikely to bias our radius measurements, as their effect is encompassed within our radius uncertainties, following the prescription in Appendix B of \citet{Leleu2021}.

\subsubsection{Radial velocity analysis}\label{subsubsect:RVs_TOI4342}

\begin{figure*}[t]
\centering
    \includegraphics[width=\textwidth]{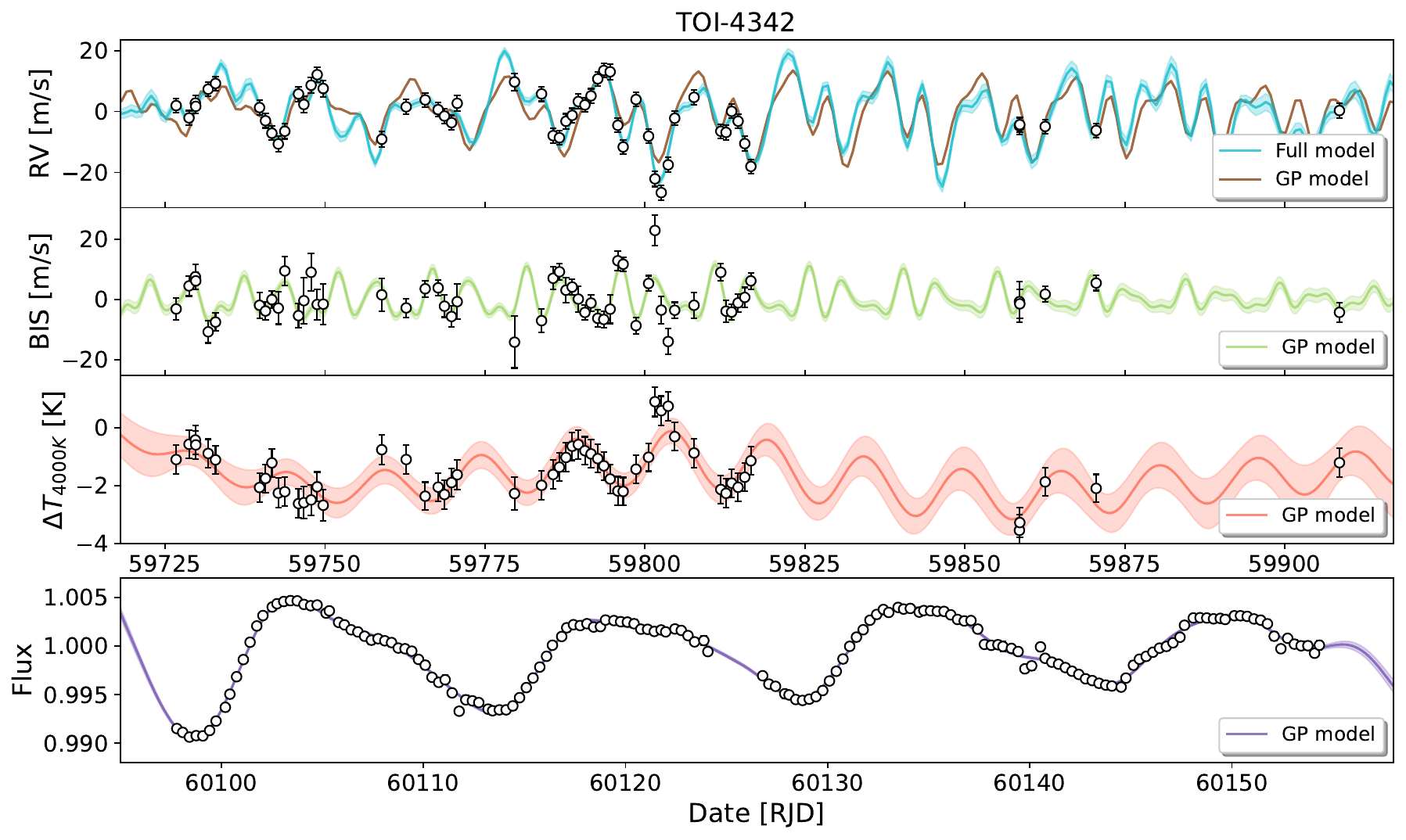}
     \caption{Time series of the RVs, FWHM, $\Delta T_{4000\,\mathrm{K}}$, and TESS binned photometry (sector 66 and 67) of TOI-4342 after correction for the inferred offsets. The measurements are shown as white dots with error bars that include the fit jitter. In the top panel, the full RV + activity model with its 1$\sigma$ uncertainty is shown in blue, while the GP-only component is shown in brown. In the three other panels, the activity models are displayed in green, pink, and purple, with their associated shaded uncertainties.}
     \label{fig:RVs_full_model_4342}
\end{figure*}

Similar as TOI-4336~A, we modeled the ESPRESSO LBL RVs of TOI-4342. We first adopted a two-planet model, assuming a uniform prior on the semi-amplitude, $K \sim \mathcal{U}(0,10)\,\mathrm{m\,s^{-1}}$, and normal priors on the orbital periods and times of conjunction based on our photometric analysis (see Sect.~\ref{subsubsect:photometry_TOI4342}). We also used a multidimensional GP framework to model the strong activity of the star (see Sect.~\ref{subsect:stellar_rot_activity}), simultaneously fitting the LBL RVs, TESS photometry, LBL $\Delta T_{4000K}$, and CCF BIS time series. 

For the GP modeling, we used the TESS SAP flux from Sectors 66 and 67, which are consecutive and temporally close to the RV observations. We binned the light curves in 8-hour intervals, effectively removing transit signals while preserving stellar variability. 

Our GP model is defined as

\begin{multline*} 
RV = A_1 \, G_1(t) \\ 
BIS = A_2 \, G_2(t) \\ 
F_\mathrm{TESS} = A_3 \, G_3(t)\\
\Delta T_{4000K} = B_3 \, G_3(t)\\ 
\end{multline*}
where $G_1(t)$, $G_2(t)$, and $G_3(t)$ are three GPs with a QP covariance kernel (Eq.~\ref{eq:qp_kernel}).
Here again, the GP hyper-parameters $P_{\mathrm{rot}}$ and $P_{\mathrm{dec}}$ are shared among all datasets, while distinct coherence scales $O_{\mathrm{amp}}$ are fit for the RVs, BIS, and photometric indicators (TESS and $\Delta T_{4000K}$). 

We used uniform priors $\mathcal{U}(10,100)$ days for the rotation period, $\mathcal{U}(20,200)$ days for the active-region decay timescale, log$\mathcal{U}(0.01,10)$ for the three coherence scales, and $\mathcal{U}(0,50)$ for the amplitudes. We also fit a jitter and offset term ($\sigma$, $\mu$) for each dataset.

After fitting this two-planet plus multi-GP model, the RV residuals exhibited excess power in the periodogram near $\sim50$ days (see middle panel Fig.~\ref{fig:TOI4342_periodograms_residuals}). This peak does not correspond to any harmonic of the rotation signal identified by the GP at $P_\mathrm{rot}=14.71^{+0.08}_{-0.07}$ days. We therefore hypothesized the presence of an additional planetary candidate and tested this by doing model comparison between the two-planet and three-planet models. For this third candidate, we adopted uniform priors of $\mathcal{U}(2,100)$ days for the orbital period and $\mathcal{U}(0.001,10)\,\mathrm{m\,s^{-1}}$ for the semi-amplitude.

We found that the three-planet model is strongly favored with a BIC of -987.78 compared to -976.39 for the two-planet model. Including the third planet also improved the semi-amplitude precision for planets b and c (see Table~\ref{tab:BIS_RVsmodel_TOI4342}). For the third candidate, we derived an orbital period of $47.5 \pm 1.3$ days, semi amplitude of 4.49 $\pm$ 0.74~$\mathrm{m~s^{-1}}$, and minimum mass of $17.8 \pm 3.0$~\me. However, the time of conjunction is poorly constrained, $T_{0,d}=58795 \pm 27$ RJD, as our RVs cover fewer than two orbital cycles. We therefore classify this signal as a planetary candidate rather than a confirmed planet. The three-planet plus multi-GP model removes any significant power in the periodogram of the residuals (see Fig.~\ref{fig:TOI4342_periodograms_residuals}, bottom panel). We also searched for transits in the available TESS light curves and were able to exclude 98.8\% of the allowed parameter space (see Sect.~\ref{subsect:tess_planet_d}). This suggests that the candidate is most likely non-transiting, although the upcoming TESS observations in 2026 should be able to confirm this scenario. 

\begin{table}[h]
\small
\caption{Model comparison RV analysis TOI-4342.}
\centering
\renewcommand{\arraystretch}{1.3}
\begin{center}
\begin{tabular}{lcccc}
\hline\hline
Model & BIC & $K_b ~(\mathrm{m\,s^{-1}})$ & $K_c ~(\mathrm{m\,s^{-1}})$ & $K_d ~(\mathrm{m\,s^{-1}})$ \\
\hline
    2P & -976.39 & 3.96$^{+0.90}_{-0.91}$ & 1.45$^{+0.80}_{-0.74}$ & - \\
    3P & -987.78 & 3.78$^{+0.65}_{-0.67}$ & 1.97$^{+0.57}_{-0.56}$ & 4.49$^{+0.73}_{-0.75}$ \\
\hline
\end{tabular}
\label{tab:BIS_RVsmodel_TOI4342}
\end{center}
\end{table}

We further tested scenarios with circular and eccentric orbits for one, two, or all planets using the parametrization of \citet{Eastman13}. None of these alternatives were statistically favored over the three-planet circular model. We therefore adopt the three-planet circular model as our final solution.

\begin{figure}[h]
\centering
    \includegraphics[width=0.48\textwidth]{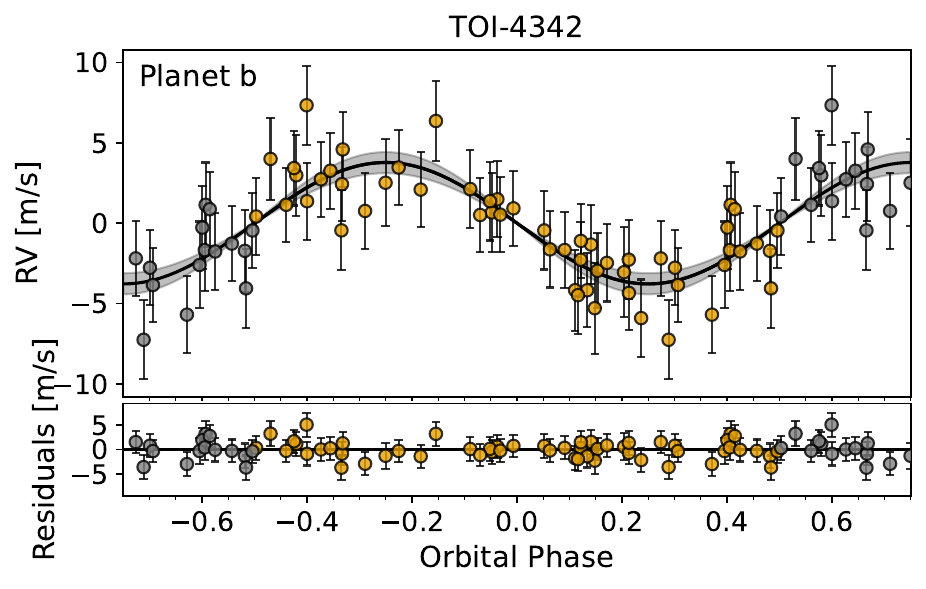}
    \includegraphics[width=0.48\textwidth]{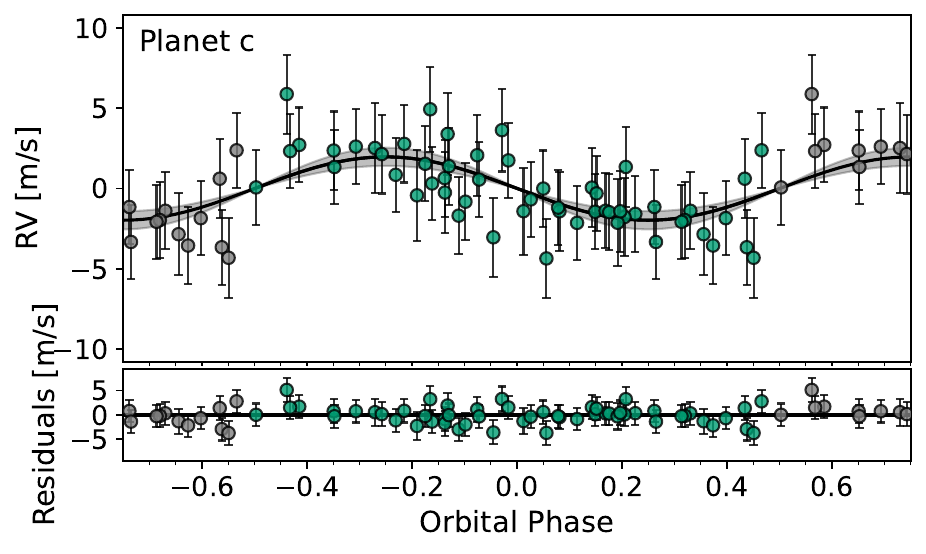}
    \includegraphics[width=0.48\textwidth]{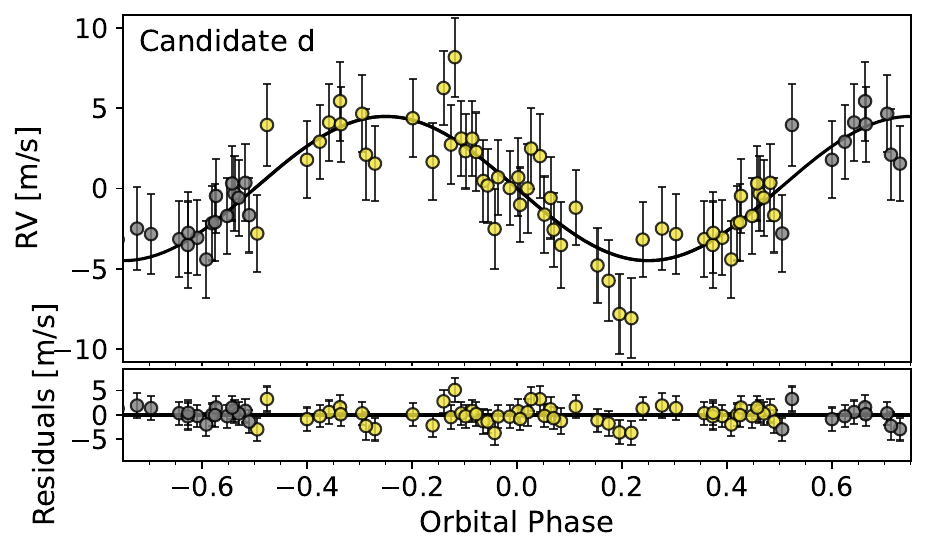}
     \caption{Phase-folded ESPRESSO RVs with the best-fit model from our analysis (Sect.~\ref{subsubsect:RVs_TOI4342}) and the corresponding residuals for TOI-4342 b (top), TOI-4342 c (middle), and TOI-4342 candidate d (bottom). The error bars include the fit jitter added in quadrature.}
     \label{fig:RVs_planets_folded_4342}
\end{figure}

The RV, BIS, $\Delta T_{4000K}$, and TESS time series, together with the best-fit model, are shown in Fig.~\ref{fig:RVs_full_model_4342}, and the phase-folded RVs of the three planets in Fig.~\ref{fig:RVs_planets_folded_4342}. The priors and posteriors of the fit are listed in Table~\ref{tab:posteriors_RV_TOI4342}, and the derived planetary parameters are summarized in Table~\ref{tab:planetaryparams_TOI4336_TOI4342}.

We confirmed the planetary natures of planet b and c by measuring their masses to be 7.3 $\pm$ 1.3~\me and 4.8 $\pm$ 1.4~\me, respectively.
We also determined the stellar rotation period to be $14.69^{+0.08}_{-0.07}$ days from our GP modeling.

\section{Discussion}\label{sect:discussion}

We report the characterization of two planetary systems around the relatively active M dwarfs TOI-4336~A and TOI-4342. TOI-4336~A hosts two transiting planets: a temperate sub-Neptune on a 16.3-day orbit with a radius of 2.11 $\pm$ 0.07~\re and a mass of 3.21 $\pm$ 0.39~\me, together with an inner super-Earth on a 7.6-day orbit with a radius of 1.20 $\pm$ 0.06~\re and a mass of 1.54 $\pm$ 0.13~\me. Assuming a null albedo and full heat redistribution, their equilibrium temperatures are of 309 $\pm$ 7 K and 400 $\pm$ 9 K, respectively. TOI-4342 has two very similar size transiting sub-Neptunes with a radius of 2.33 $\pm$ 0.09~\re for planet b and 2.35 $\pm$ 0.09~\re for planet c. However, their masses seem to be slightly different, 7.2 $\pm$ 1.3~\me for planet b compared to 4.8 $\pm$ 1.4~\me for planet c, still consistent within 1-$\sigma$. They orbit with periods of 5.54 and 10.69 days, close to a 2:1 period ratio, but we do not detect any evidence for significant TTVs (see Sect.~\ref{subsubsect:photometry_TOI4342}). We calculate their equilibrium temperature to be 639 $\pm$ 20 K and 514 $\pm$ 16 K, respectively. We also detected in the RVs an additional planet candidate orbiting on a 47.7-day orbit with a minimum mass of 17.8 $\pm$ 3.0~\me, likely a Neptune-like planet, more massive than its inner companions. However, with less than 2 orbital period covered from our observations, further confirmation is needed for this planetary candidate.

\subsection{Low density sub-Neptunes around M dwarfs}

\begin{figure}[t]
  \centering
    \includegraphics[width=0.48\textwidth]{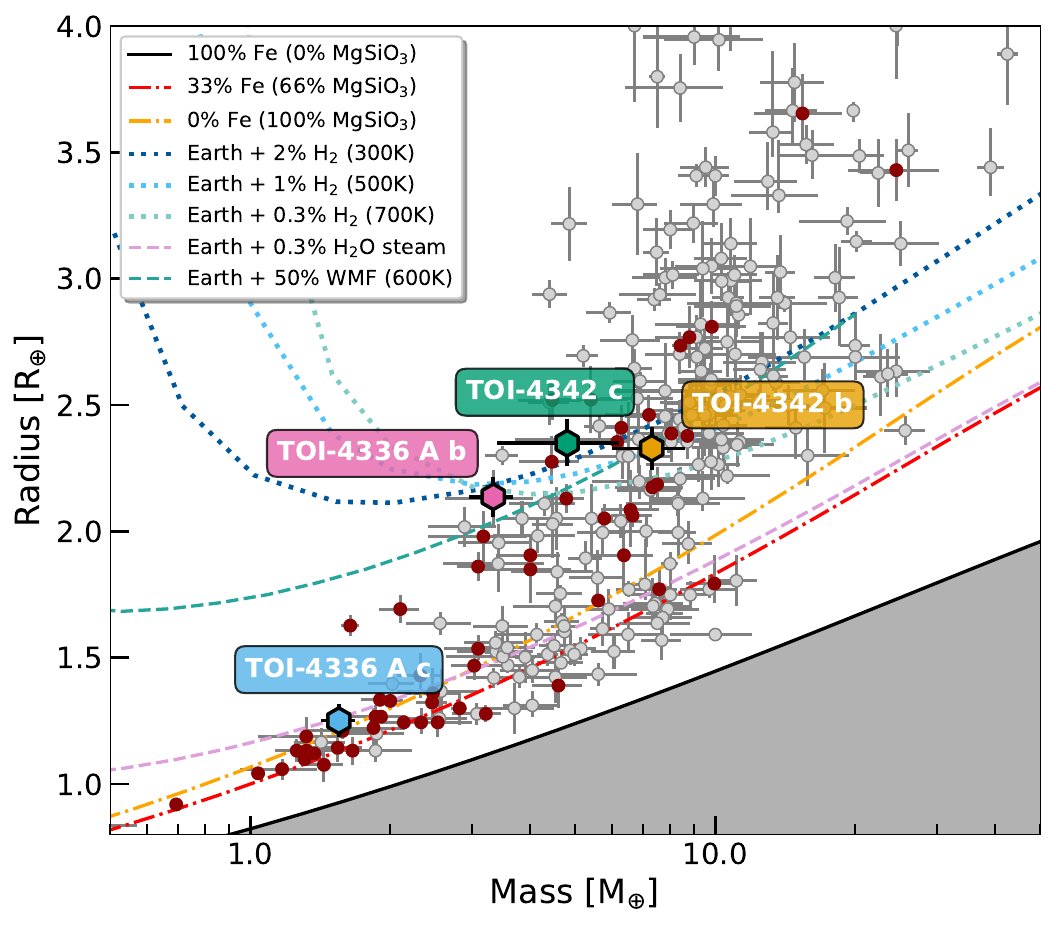}
  \caption{Mass-radius diagram of small exoplanets (with radii ranging from 1–4~\re) with precise densities from the PlanetS catalog. The red (gray) dots correspond to exoplanets orbiting M dwarfs (FGK dwarfs). The composition lines of pure silicates (yellow dashed), Earth-like planets (red dashed), and pure iron (solid black) from \citet{Zeng2019} are displayed. The four transiting planets of this study are highlighted as hexagons with their name. Composition lines of Earth with an water steam atmospheres from \citet{Turbet2020} and from \citet{Aguichine2025} models are represented in pink dashed line and dark turquoise dashed line, respectively. Compositions of Earth with an hydrogen-rich atmospheres from \citet{Zeng2019} are represented in dotted lines. They all are adapted with the equilibrium of the related planet.
  This plot has been generated with \texttt{mr-plotter} \citep[][ \url{https://github.com/castro-gzlz/mr-plotter/}]{MR-plotter2024}.}
  \label{fig:MR-diagram}
\end{figure}

TOI-4336~A~b, TOI-4342~b, and TOI-4342~c expand the current population of characterized sub-Neptunes ($2~R_\oplus < R_p < 4~R_\oplus$) orbiting M dwarfs. Figure~\ref{fig:MR-diagram} shows the mass--radius distribution of small planets from the PlanetS catalog \citep{Otegi2020, Parc2024}, with planets orbiting M dwarfs highlighted in red. With bulk densities of 1.87$^{+0.31}_{-0.28}$, 3.18$^{+0.70}_{-0.63}$, and 2.01$^{+0.65}_{-0.60}$~g~cm$^{-3}$, respectively, TOI-4336~A~b, TOI-4342~b, and TOI-4342~c follow the trend reported by \citet{Parc2024, Parc2025}, who found that small sub-Neptunes ($1.8~R_\oplus < R_p < 2.9~R_\oplus$) around M dwarfs tend to be less dense than their counterparts orbiting FGK stars.

Using the methodology of \citet{Parc2025}, we performed a Mann--Whitney U test \citep{Wilcoxon1945,Mann1947} on 10\,000 bootstrap realizations drawn from the density distributions of each planet to evaluate how the inclusion of these new planets affects the inferred density contrast between M-dwarf and FGK-dwarf systems. The up-to-date PlanetS catalog (as of 24 November 2025) for this population currently contains 22 planets around M dwarfs and 84 around FGK stars, after removal of the TTVs characterized planets to isolate any effect from resonance-state on the density found in \citet{Leleu2024}. Excluding TOI-4342~c, which does not meet the mass-precision criterion for inclusion in the PlanetS catalog, we obtain a median p-value of 0.006, with 95.5\% of the realizations rejecting the null hypothesis (p-value $<$ 0.05), consistent with \citet{Parc2025}. 

We investigate the possible compositions of the four planets by comparing their inferred masses and radii to theoretical composition curves from various models displayed in Fig.~\ref{fig:MR-diagram}, before performing a more detailed interior analysis in Sect.~\ref{subsect:interior_model}. 

TOI-4336~A~c is the only super-Earth of the systems studied in this paper. With a density of 4.35$^{+0.86}_{-0.71}$~g~cm$^{-3}$ and an equilibrium temperature of $\sim$400~K, it lies above the Earth-like composition line, on the pure-silicate-core model. Its density may be explained by a smaller iron fraction in its core, or by the presence of volatiles, such as water vapor, in its atmosphere, as suggested by the composition line of \citet{Turbet2020} (pink dashed line). 

Among the three sub-Neptunes, TOI-4336~A~b has a very low density of 1.87$^{+0.31}_{-0.28}$~g~cm$^{-3}$ and an equilibrium temperature of $\sim$309~K. At this temperature, water could be present either in a condensed phase or as steam \citep{Venturini2024}. Yet, given its low density, the planet lies above the pure-water envelope cases, suggesting that some amount of hydrogen and helium is likely required in its atmosphere, either predominantly H/He or possibly mixed with water, to account for its bulk properties. For instance, it is consistent with an Earth-like core enveloped by a 2\% H$_2$ atmosphere from \citet{Zeng2019} models (dark blue dotted line). 

TOI-4342~b and TOI-4342~c receive higher irradiation and are denser than TOI-4336~A~b. Their measured densities (3.18$^{+0.70}_{-0.63}$ and 2.01$^{+0.65}_{-0.60}$~g~cm$^{-3}$) together with their equilibrium temperatures ($\sim$639 and $\sim$514~K) allow for different interior structure interpretations. TOI-4342~b is compatible with an Earth-like core surrounded by a 50\% steam atmosphere (dark turquoise dashed line) when accounting for contraction and cooling effects \citep{Aguichine2025}, while it lies slightly above the model with a thin ($\sim$0.3\%) H$_2$ envelope (light turquoise dotted line), a combination of both components also remains possible. TOI-4342~c, although within $1\sigma$ of the mass of TOI-4342~b, is less irradiated and could be less dense. Here again, its properties likely point toward the presence of some amount of hydrogen and helium, as it lies above the pure-water envelope cases. For example, the 1\% H$_2$ envelope model from \citet{Zeng2019} lies about $2\sigma$ below its measured density, implying either a larger H/He mass fraction or a mixture of supercritical H$_2$O and H/He. 

In summary, only TOI-4342~b could plausibly be a steam-rich water world, whereas TOI-4336~A~b and TOI-4342~c likely require a significant hydrogen--helium component to reproduce their densities. This highlights the important role of equilibrium temperature in shaping the possible compositions of the two TOI-4342 planets, despite their nearly identical masses and radii. TOI-4336~A~c is a slightly low-density super-Earth that could either be iron-depleted or volatile-rich.

\subsection{Detailed interior modeling}\label{subsect:interior_model}

We analyzed the potential composition of the planets in more detail using an inference model based on \citet{Dorn2015} with updates in \citep{DeWringer}. The inference uses a physical forward model based on \citet{dorn_generalized_2017} with updates described in \citet{dorn2021hidden, luo_majority_2024}.
The underlying forward model consists of three layers: an iron core, a silicate mantle, and a H$_2$-He-H$_2$O or a pure H$_2$O steam atmosphere and is described in more detail in Appendix \ref{app:interior}.

For the three sub-Neptunes, we consider a H$_2$-He-H$_2$O atmosphere, while for the super-Earth case (TOI-4336~A~c), we consider the possibility of water but not the presence of a primordial gas envelope. 
Water can be incorporated in the planet’s mantle, core melts, and outer layers, affecting densities and melting temperatures, with phase-dependent behavior modeled following recent EOS and solubility studies \citep{dorn_hidden_2021, luo_majority_2024, bajgain_structure_2015, haldemann_aqua_2020}.

The data considered include planetary mass, radius, equilibrium temperature, and stellar abundances (Fe/H, Mg/H, Si/H, derived in Sect.~\ref{subsect:spectro_stellar}); except for sub-Neptunes, where the stellar abundances are used to inform a predefined silicate-to-iron ratio. Also specific to sub-Neptunes, we consider undifferentiated interiors \citep{young2025differentiation}. The priors of the inference can be found in Table~\ref{tab:priors_inference}.

The results of the inference model are summarized in Table~\ref{tab:inferenceresult}, Fig.~\ref{fig:int_post_4336bc}, and \ref{fig:int_post_4342bc}.
We find that based on the stellar proxy on the bulk Fe/Si-ratio and the mass-radius data, the super-Earth TOI-4336~A~c can be explained by a purely rocky composition with a core mass fraction (CMF) of $\sim0.31$, which corresponds to a roughly Earth-like CMF. 
Nevertheless, the mode of the posterior distribution is lower than the mean in the case of the core mass, which means that lower CMF are plausible in line with Fig.~\ref{fig:MR-diagram}. 
However, TOI-4336~A~c is also consistent with a rocky planet that contains $\sim 2\%$ of water. 
The high core-to-mantle mass ratio of almost 1 requires a higher water mass fraction than the composition line shown in Fig.~\ref{fig:MR-diagram} to match the observed bulk density. 
 
We find that all three sub-Neptunes show super-solar metallicity, which increases the required atmosphere mass fractions compared to the composition lines shown in Fig.~\ref{fig:MR-diagram}.
As expected, the atmosphere mass fraction scales with the bulk density of the planet.
For the very low density planet TOI-4336~A~b, we find an atmosphere mass fraction (AMF) of $\sim 3.7\%$, while the for TOI-4342~c we find an AMF of $\sim 2.9\%$ and for TOI-4342~b we find and AMF of $\sim 1.8\%$. 
However, the exact atmosphere mass and metallicity remain unconstrained for all three sub-Neptunes.
Follow-up observations in order to characterize the atmosphere of these planets are needed to break the degeneracy between mass and metallicity and better constrain the composition of these planets.

\begin{table}[h]
\centering
\small
\caption{Inference results for the internal compositions of TOI-4342\,b, TOI-4342\,c, TOI-4336~A\,b, and TOI-4336~A\,c. }\label{tab:inferenceresult}
\renewcommand{\arraystretch}{1.1}
\begin{tabular}{lccc}
    \hline\hline
     \textbf{Parameters}  &  & \textbf{Values} & \\
      &   TOI-4342\,b & TOI-4342\,c & TOI-4336~A\,b\\
    \hline \\[-6pt]
    $M_\mathrm{atm} \: (M_\oplus)$ &  $0.13^{+0.17}_{-0.07  } $ & $0.15^{+0.16}_{-0.08}$ & $0.12^{+0.13}_{-0.06}$\\
    $M_\mathrm{core+mantle} \: (M_\oplus)$ &  $7.29^{+1.21}_{-1.18}$ & $5.03^{+1.28}_{-1.31}$ &  $3.15^{+0.37}_{-0.38}$\\
    $Z_\mathrm{env}$ &  $0.42 ^{+0.26}_{-0.24} $ & $0.31^{+0.24}_{-0.19}$ & $0.38^{+0.21}_{-0.21}$\\[+3pt]
    \hline \\[-7pt]
     &  TOI-4336~A\,c & including water & purely rocky \\
    \hline \\[-6pt]
    $M_\mathrm{water} \: (M_\oplus)$ & &$0.03^{+0.03}_{-0.02} $ & - \\
    $M_\mathrm{mantle} \: (M_\oplus)$ &  &$0.75^{+0.24}_{-0.20}$ & $1.12^{+0.23}_{-0.26}$\\
    $M_\mathrm{core} \: (M_\oplus)$ &  & $0.74^{+0.24}_{-0.26}$ & $0.50^{+0.33}_{-0.23}$\\
    \bottomrule
\end{tabular}
\tablefoot{$Z_\mathrm{env}$ corresponds to the envelope metal (here water) mass fraction.}
\end{table}

\subsection{Prospects for atmospheric characterization}

\begin{figure}[t]
  \centering
    \includegraphics[width=0.48\textwidth]{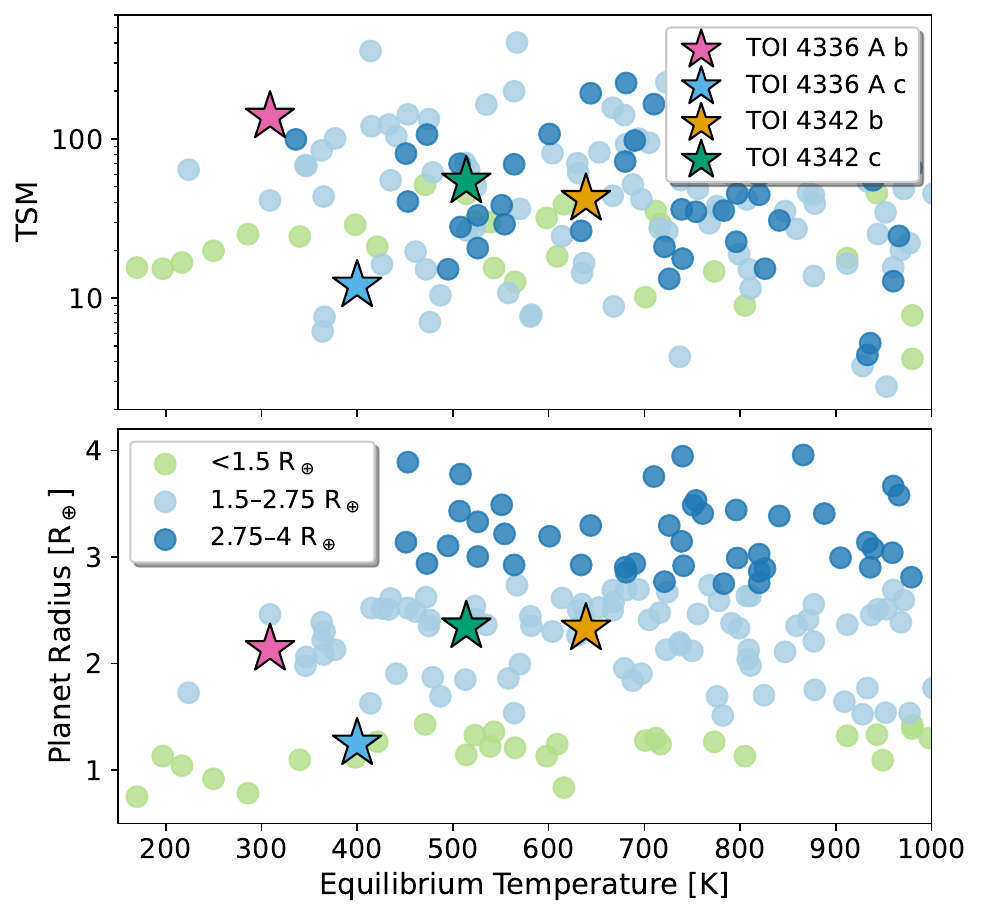}
  \caption{Transmission spectroscopy metric (top), and planetary radius (bottom) as a function of the equilibrium temperature for small ($<4$~\re) well-characterized planets from the PlanetS catalog \citep{Otegi2020,Parc2024}. The population is divided following \citet{Kempton2018}: terrestrial planets ($<1.5$~\re) in light green, small sub-Neptunes ($1.5$–$2.75$~\re) in light blue, and large sub-Neptunes ($2.75$–$4$~\re) in dark blue. TOI-4336~A~b, TOI-4336~A~c, TOI-4342~b, and TOI-4342~c are highlighted as pink, blue, orange, and green stars, respectively.}
  \label{fig:TSM_Rp_Teq}
\end{figure}

To assess the atmospheric characterization potential of our four transiting planets, we computed their transmission spectroscopy metric and emission spectroscopy metric \citep[TSM and ESM;][]{Kempton2018} and compared them to the population of small ($<4$~\re), well-characterized planets in the PlanetS catalog \citep{Otegi2020, Parc2024}; see Fig.~\ref{fig:TSM_Rp_Teq}. We obtained TSM values of 138, 12, 42, and 54 and ESM values of 0.82, 1.01, 3.86, and 1.95 for TOI-4336~A~b, TOI-4336~A~c, TOI-4342~b, and TOI-4342~c, respectively.

TOI-4336~A~b immediately stands out. With a high TSM of 138 and a relatively low equilibrium temperature of 309~K, it ranks among the most promising atmospheric targets in this temperature regime. It is, in fact, the highest-TSM planet with $T_{\mathrm{eq}}<320$~K, alongside benchmark objects such as K2-18~b ($T_{\mathrm{eq}}=309$~K, TSM=41; \citealt{Howard2025}) and LHS~1140~b ($T_{\mathrm{eq}}=224$~K, TSM=64; \citealt{Cadieux2024a}), both considered top-priority JWST targets. In fact, the planet has already been observed with HST (Cycle 29 GO16875, PI: Dransfield) and JWST (Cycle 3 GO4711, PI: Hu). Given its favorable properties, this planet is likely to become one of the best-studied members of its category in the near future.

TOI-4336~A~c is the least irradiated super-Earths known in the PlanetS catalog, aside from the TRAPPIST-1 system. It occupies a similar regime to LHS~1140~c ($T_{\mathrm{eq}}=421$~K, TSM=21; \citealt{Cadieux2024a}). As with LHS~1140, the TOI-4336 system is thus particularly well suited for comparative planetology, enabling atmospheric studies of two planets orbiting the same star. With a TSM of 12, TOI-4336~A~c falls within the “top quintile” category for terrestrial planets defined by \citet{Kempton2018} and lies well above the updated follow-up threshold of $\sim$4.5 recommended by \citet{Guerrero2021} for the 37 most atmospherically characterizable terrestrial planets. Additionally, with an escape velocity of 12.5~km~s$^{-1}$ and a relative cumulative XUV irradiation of 63.5 computed following \citet{Zahnle2017}, the planet lies very close to the “cosmic shoreline.” It therefore stands out as a particularly promising candidate for emission spectroscopy, with an ESM of 1.01, similar to targets included in the Rocky Worlds JWST DDT program\footnote{\url{https://rockyworlds.stsci.edu/index.html}}.

TOI-4342~b and TOI-4342~c, with TSM values of 42 and 54, also emerge as compelling candidates for atmospheric follow-up. Both lie above the median TSM of the catalog and exceed the threshold of $\sim$25 proposed by \citet{Guerrero2021} for the 100 best small sub-Neptunes. Moreover, TOI-4342 offers the added advantage of hosting two sub-Neptunes amenable to atmospheric studies within the same system. Notably, \citet{Tey2023} already highlighted TOI-4342 as one of the top ten M-dwarf multiplanet systems best suited for atmospheric characterization. Improved mass constraints, particularly for TOI-4342~c, would further enhance their suitability by enabling the $\sim$20\% mass precision required for robust interpretation of transmission spectra \citep{Batalha2019}.

\subsection{Dynamical stability analysis of the TOI-4342 system}

We analyzed the dynamical stability of the TOI-4342 system, consisting of the two transiting planets close to 2:1 mean-motion resonance and the third 47-day candidate, to assess whether this additional companion could coexist in a stable configuration with the two inner planets. To this end, we performed a suite of 100 $N$-body simulations using the \texttt{REBOUND}\footnote{\url{https://github.com/hannorein/rebound}} Python package \citep{rebound} with the \texttt{IAS15} adaptive integrator \citep{reboundias15}.

For each simulation, we initialized the three planets by drawing their orbital parameters and masses from the posterior distributions derived in Sect.~\ref{subsect:analysis_TOI4342}, in order to account for the uncertainties in the orbital solutions. This choice was made especially to explore the unconstrained time of conjunction of the candidate planet~d  that could lead to potential instabilities. As this latter most-likely does not transit, we assumed it to be coplanar with planet~c. Each simulation was integrated for 1~Myr, with planetary parameters recorded every 1000~years.

For all 100 simulations, we find that all three planets remain dynamically stable over the entire 1~Myr timespan. The orbital parameters remain bounded and nearly constant, and the eccentricities stay below 0.02. We observe clear dynamical interactions between planets~b and~c, which lie close to a 2:1 mean-motion resonance, whereas the candidate~d appears too distant to interact significantly with the inner pair. 

In addition, we compute the planet spacing as defined in \citet{Pu2015} (Eq. 1 and 2) between planet b and c and between planet c and candidate d. We find values of planet spacing of K=16 and K=27, respectively, both larger than the critical spacing of $\sim 12$, suggesting that the system should be stable on a billion-year timescale.

\section{Conclusions}\label{sect:conclusions}
 
We presented the characterization of two planetary systems orbiting the M dwarfs TOI-4336~A and TOI-4342. Each system hosts two transiting planets previously validated by \citet{Timmermans2024,Timmermans2026} and \citet{Tey2023}. For TOI-4342, we updated the photometric analysis of the TESS and LCO light curves by incorporating four additional TESS sectors, thereby refining the ephemeris and transit parameters. Using high-resolution ESPRESSO and NIRPS spectroscopy, we performed a detailed stellar characterization of both host stars. By extracting RVs with the LBL method \citep{Artigau2022}, we significantly improved the RV precision for both targets. Joint modeling of TESS photometry, RV indicators, and RV time series within a Gaussian-process framework enabled us to constrain the stellar rotation periods to $35.5 \pm 1.1$ days for TOI-4336~A and $14.69 \pm 0.08$ days for TOI-4342.

We confirmed the planetary nature of the four transiting planets by precisely measuring their masses. TOI-4336~A~b is a sub-Neptune with a radius of $2.14 \pm 0.08$~\re{} and a mass of $3.33 \pm 0.40$~\me; TOI-4336~A~c is a super-Earth with $1.25 \pm 0.07$~\re{} and $1.55 \pm 0.13$~\me. TOI-4342~b and TOI-4342~c are both sub-Neptunes, with radii of $2.33 \pm 0.09$~\re{} and $2.35 \pm 0.09$~\re{} and masses of $7.3 \pm 1.3$~\me{} and $4.8 \pm 1.4$~\me, respectively. In addition, we identified a likely non-transiting candidate planet around TOI-4342 with a period of $47.5 \pm 1.3$ days and a minimum mass of $17.8 \pm 3.0$~\me. The current TESS coverage excludes transits with a probability of 98.8\%. Dynamical analyses indicate that the system remains stable in the presence of this additional planet. While our current RV time baseline is insufficient to fully constrain its orbital phase, future observations will allow confirmation and ephemeris determination. 

One of the original motivations of this ESPRESSO program was to investigate differences in the density distribution of small sub-Neptunes around M dwarfs compared to FGK stars. We derived the densities of TOI-4336~A~b, TOI-4336~A~c, TOI-4342~b, and TOI-4342~c to be $1.87 \pm 0.30$, $4.35 \pm 0.79$, $3.18 \pm 0.67$, and $2.01 \pm 0.63$~g~cm$^{-3}$. The sub-Neptunes densities' follow the trend reported by \citet{Parc2024,Parc2025}, further supporting the conclusion that these two populations do not share the same underlying density distribution, with planets around M dwarfs being less dense than their counterparts around earlier-type stars. This strengthens the case for distinct formation or evolutionary pathways for small sub-Neptunes around low-mass stars.

We also explored the possible compositions of the planets using mass–radius diagram and a Bayesian inference framework. Using stellar abundances derived with the NIRPS spectra, we find that the super-Earth TOI-4336~A~c has a CMF $\sim$ 0.31, close to that of Earth, but could also have $\sim$ 2\% of water in its interior or/and atmosphere. The sub-Neptunes show super-solar metallicity and require H$_2$--He--H$_2$O envelopes to match their observed bulk densities, with AMF of $\sim$3.7\%, $\sim$2.9\%, and $\sim$1.8\% for TOI-4336~A~b, TOI-4342~c, and TOI-4342~b, respectively.

Finally, we evaluated the prospects for atmospheric characterization. TOI-4336~A~b stands out as one of the most favorable known targets for transmission spectroscopy. More broadly, we find that both TOI-4336~A and TOI-4342 offer benchmark systems for exploring the diversity in sub-Neptune atmospheres composition with JWST. In particular, the recent transmission spectroscopy programs on TOI-4336~A~b will greatly benefit from the precise mass measurement provided in this study, which is essential for a robust interpretation of its atmospheric spectrum.

Overall, the TOI-4336~A and TOI-4342 systems provide valuable additions to the growing sample of well-characterized M-dwarf systems. With precisely measured planetary masses, radii, densities, stellar rotation periods, and strong potential for atmospheric follow-up, they will serve as benchmark laboratories for studying the formation, composition, and evolution of small planets around low-mass stars. \\

\section*{Data availability}
The data used in this study can be downloaded and visualized on the DACE platform: \url{https://dace.unige.ch/openData/?record=10.82180/dace-pwgeam18}.

\begin{acknowledgements}
We thank the anonymous referee for their valuable comments, which helped improve the manuscript. 
This work has been carried out within the framework of the NCCR PlanetS supported by the Swiss National Science Foundation under grants 51NF40\_182901 and 51NF40\_205606. \\
C.D. acknowledges support from the Swiss National Science Foundation under grant TMSGI2\_211313.\\
ED-M acknowledges the support by the Ram\'on y Cajal contract RyC2022-035854-I funded by MICIU/AEI/\url{10.13039/501100011033} and by ESF+. \\
N.A-D. acknowledges the support of FONDECYT project 1240916. \\
KAC acknowledges support from the TESS mission via subaward s3449 from MIT.\\
The Board of Observational and Instrumental Astronomy (NAOS) at the Federal University of Rio Grande do Norte research's activities are supported by continuous grants from the Brazilian funding agency CNPq. JRM acknowledges CNPq research fellowships (Grant No. 308928/2019-9).\\
XDe acknowledges funding from the French ANR under contract number ANR\-24\-CE49\-3397 (ORVET), and the French National Research Agency in the framework of the Investissements d'Avenir program (ANR-15-IDEX-02), through the funding of the ``Origin of Life'' project of the Grenoble-Alpes University. \\
XDu acknowledges the support from the European Research Council (ERC) under the European Union’s Horizon 2020 research and innovation programme (grant agreement SCORE No 851555) and from the Swiss National Science Foundation under the grant SPECTRE (No 200021\_215200). \\
JIGH acknowledges financial support from the Spanish Ministry of Science, Innovation and Universities (MICIU) projects PID2020-117493GB-I00 and PID2023-149982NB-I00. \\
We acknowledge grants Spanish program Unidad de Excelencia Mar\'ia de Maeztu CEX2020-001058-M and 2021-SGR-1526 (Generalitat de Catalunya).\\
NCS is co-funded by the European Union (ERC, FIERCE, 101052347). Views and opinions expressed are however those of the author(s) only and do not necessarily reflect those of the European Union or the European Research Council. Neither the European Union nor the granting authority can be held responsible for them. This work was supported by FCT -- Funda\c{c}\~ao para a Ci\^encia e a Tecnologia through national funds by these grants: UIDB/04434/2020 \url{https://doi.org/10.54499/UIDB/04434/2020}, UIDP/04434/2020 \url{https://doi.org/10.54499/UIDP/04434/2020}. \\
JV acknowledges support from the Swiss National Science Foundation (SNSF) under grant PZ00P2\_208945. \\
Funding for the TESS mission is provided by NASA's Science Mission Directorate.  We acknowledge the use of public TESS data from pipelines at the TESS Science Office and at the TESS Science Processing Operations Center. Resources supporting this work were provided by the NASA High-End Computing (HEC) Program through the NASA Advanced Supercomputing (NAS) Division at Ames Research Center for the production of the SPOC data products. This paper includes data collected by the TESS mission that are publicly available from the Mikulski Archive for Space Telescopes (MAST). \\
This work makes use of observations from the LCOGT network. Part of the LCOGT telescope time was granted by NOIRLab through the Mid-Scale Innovations Program (MSIP). MSIP is funded by NSF. \\
This research has made use of the Exoplanet Follow-up Observation Program (ExoFOP; \url{https://doi.org/10.26134/ExoFOP5}) website, which is operated by the California Institute of Technology, under contract with the National Aeronautics and Space Administration under the Exoplanet Exploration Program. \\
\end{acknowledgements}

\bibliographystyle{aa} 
\bibliography{bib}

\clearpage
\onecolumn

\begin{appendix}

\section{Stellar parameters}

\begin{table}[h]
\small
\centering
\caption{Stellar abundances measured with NIRPS.}
\makebox[\linewidth]{
\begin{tabular}{lcc|cc}
\hline
\hline
 & \multicolumn{2}{c}{\textbf{TOI-4336 A}} & \multicolumn{2}{c}{\textbf{TOI-4342}} \\
\hline
Element & [X/H]$^*$  & Num. of lines & [X/H]$^*$  & Num. of lines \\
\hline
Fe I & $0.14\pm0.13$ & 7    & $0.26\pm0.13$  & 13\\
Mg I & $0.25\pm0.36$ & 4    & $0.34\pm0.18$  & 5\\
Si I & $-0.13\pm0.17$ & 1   & $0.52\pm0.30$  & 4\\
Ca I & $-0.20\pm0.17$ & 1   & $-0.05\pm0.15$ & 5\\
Ti I & $0.24\pm0.38$ & 7    & $-0.03\pm0.10$ & 8\\
Al I & $0.14\pm0.30$ & 2    & $0.17\pm0.14$  & 1\\
Na I & $-0.06\pm0.18$ & 1   & -- & --\\
C I  & $-0.01\pm0.24$ & 2   & -- & --\\
K I  &  $-0.14\pm0.41$ & 2  & $0.02\pm0.13$  & 2\\
OH   & $-0.15\pm0.14$ & 11  & $-0.60\pm0.09$ & 40\\[0.05cm]
\hline
[M/H] & $0.01\pm0.06$ & --  & $0.08\pm0.11$  & --\\[0.05cm]
\hline
\label{table:abundances_TOI4336TOI4342}
\end{tabular}}
\tablefoot{$^*$Molar relative-to-solar abundances}
\end{table}

\begin{table}[h]
\small
\caption{Stellar parameters for TOI-4336 A and TOI-4342.}
\centering
\renewcommand{\arraystretch}{1.2}
\makebox[\linewidth]{
\begin{tabular}{lccc}
\hline\hline
  & \textbf{TOI-4336 A}  & \textbf{TOI-4342} & Reference \\
    \hline
    \textbf{Identifiers}& &&\\
    TIC ID  & 166184428 & 354944123 & TICv8\\
    2MASS ID & J13442546-4020155 &  J21373286-7758435 & 2MASS\\
    Gaia ID & 6113245033656232448 &  6355915466181029376& Gaia DR3\\
    \hline
    \textbf{Astrometric parameters} & &\\
    Right ascension (J2000),  $\alpha$ & 13$^{\mathrm{h}}$ 44$^{\mathrm{m}}$ 25.69$^{\mathrm{s}}$ & 21$^{\mathrm{h}}$ 37$^{\mathrm{m}}$ 33.48$^{\mathrm{s}}$&Gaia DR3\\
    Declination (J2000), $\delta$ & -40$^{\circ}$ 20$^{\prime}$ 14.43$^{\prime \prime}$& -77$^{\circ}$ 58$^{\prime}$ 44.97$^{\prime \prime}$&Gaia DR3\\
    Parallax (mas) & 44.53 $\pm$ 0.04 & 16.25 $\pm$ 0.02 &Gaia DR3\\
    Distance (pc) & 22.47$\pm$ 0.02 &  61.54 $\pm$ 0.07 &Gaia DR3\\
    $\mu_{\rm{R.A.}}$ (mas yr$^{-1}$) &151.81 $\pm$ 0.03 & 120.33 $\pm$ 0.02 &Gaia DR3\\
    $\mu_{\rm{Dec}}$ (mas yr$^{-1}$) &68.40 $\pm$ 0.03 & -91.50 $\pm$ 0.02 &Gaia DR3\\
    $V_\mathrm{syst}$ (km s$^{-1}$) &18.38 $\pm$ 0.36 & -4.43 $\pm$ 0.36 & Gaia DR3\\
    \hline
    \textbf{Photometric parameters} & &\\
    TESS (mag) & 11.020 $\pm$ 0.007 & 11.032 $\pm$ 0.007 & TICv8\\
    \textit{V} (mag) &13.600 $\pm$ 0.092 & 12.669 $\pm$ 0.057 & TICv8\\
    \textit{G} (mag) &12.2569 $\pm$ 0.0005 & 11.9822 $\pm$ 0.0004 & Gaia DR3\\
    \textit{J} (mag) & 9.453 $\pm$ 0.024 & 9.832 $\pm$ 0.024 & 2MASS\\
    \textit{H} (mag) & 8.867 $\pm$ 0.046 & 9.179 $\pm$ 0.024 & 2MASS\\
    $K_s$ (mag) & 8.632 $\pm$ 0.024 & 9.018 $\pm$ 0.021 & 2MASS\\
    \hline
    \textbf{Bulk parameters} & & &\\
    Spectral type & M3.5V & M0V & This work \\
    \teff\,(K) &  3307 $\pm$ 94 & 3866 $\pm$ 92 & This work \\
    $R_\star$  (\rsol)  &  0.326 $\pm$ 0.010 & 0.598 $\pm$ 0.018 & This work \\
    $M_\star$  (\msol)  &  0.306 $\pm$ 0.008  & 0.587 $\pm$ 0.013 & This work \\
    $\rho_\star$  (\gccc)  &  12.45$^{+1.25}_{-1.12}$  & 3.86$^{+0.38}_{-0.34}$ & This work \\
    $L_\star$  (\lsol)  &  0.011 $\pm$ 0.001  & 0.075 $\pm$ 0.007 & This work\\
    \feh (dex) &  0.14 $\pm$ 0.13 & 0.26 $\pm$ 0.13 & This work\\
    \mh (dex) &  0.01 $\pm$ 0.06  & 0.08 $\pm$ 0.11 & This work\\
    $\log g_*$ (cm\,s$^{-2}$) &  4.897 $\pm$ 0.029  & 4.652 $\pm$ 0.028 & This work\\
    $P_\mathrm{rot}$ (days) &  35.5 $\pm$ 1.1  & $14.691^{+0.080}_{-0.070}$ & This work\\
   \hline
\end{tabular}}
\tablebib{(TICv8) \citet{Stassun2019}; (2MASS) \citet{2MASS2006}; (Gaia DR3) \citet{Gaia2018}}
\label{tab:Stellar parameters}
\end{table}

\clearpage
\section{Data analysis}

\begin{figure}[h]
    \centering
    \includegraphics[width=0.48\textwidth]{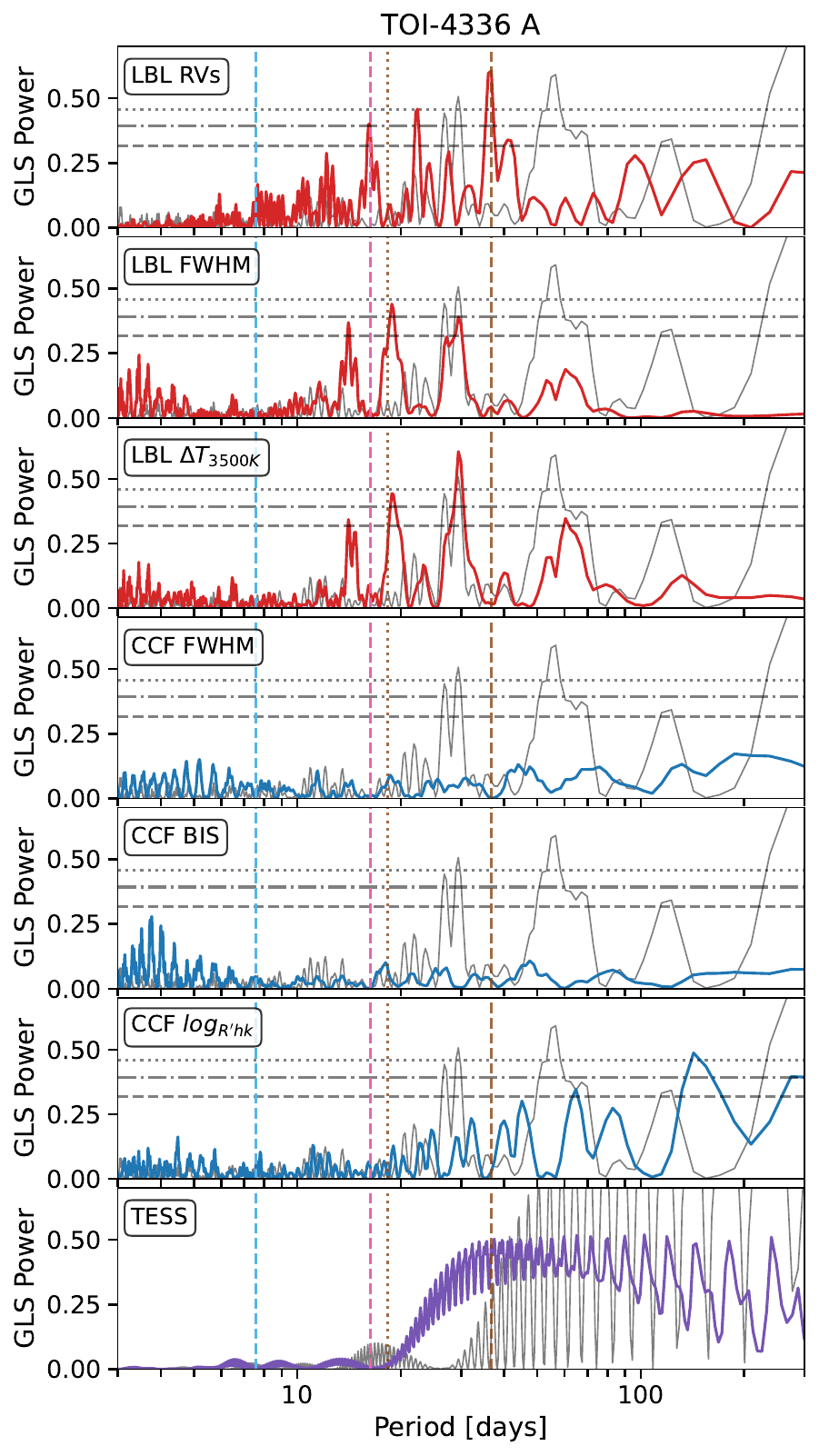}
    \includegraphics[width=0.48\textwidth]{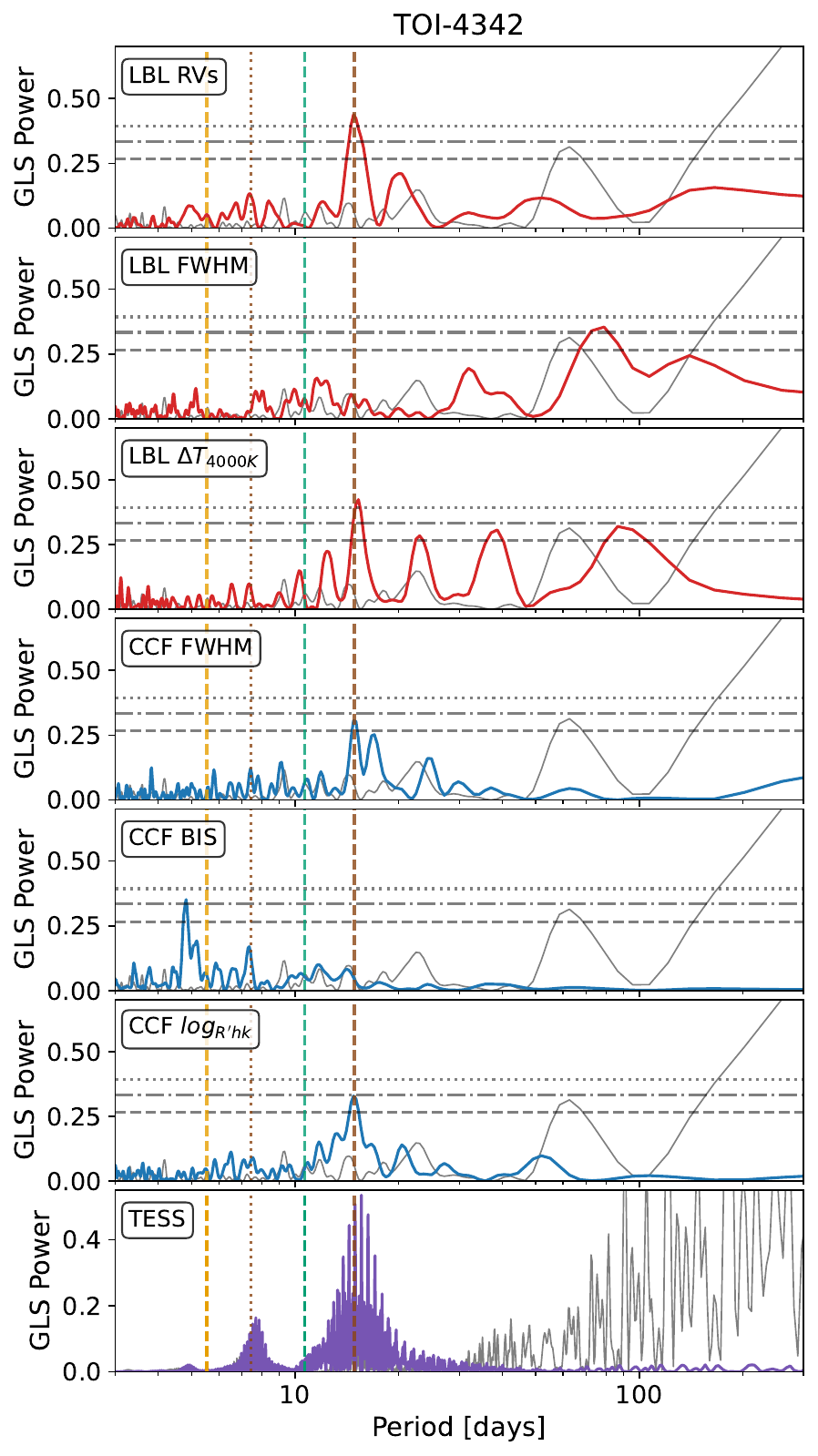}
    \caption{GLS periodograms of the ESPRESSO RVs and activity indicators for TOI-4336~A (left) and TOI-4342 (right) obtained with the LBL (red) and CCF (blue) methods, and of TESS photometry (purple). The gray curve corresponds to the window function. The vertical brown lines indicate the possible stellar rotation period at $\sim$36 days (resp $\sim$14 days) (dashed line) and its first harmonic at $\sim$18 days (resp $\sim$7 days) (dotted line). The pink and blue (resp orange and green) vertical dashed lines mark the expected orbital periods of planets b and c, respectively. The 10\%, 1\%, and 0.1\% False Alarm Probability (FAP) levels are shown as gray horizontal dashed, dash-dotted, and dotted lines, respectively.} 
    \label{fig:TOI4336_periodograms}
\end{figure}

\begin{table*}[h]
\tiny
\caption{Median values and 68$\%$ confidence intervals of the posterior distributions of the RV fit of TOI-4336 A.}
\centering
\renewcommand{\arraystretch}{1.1}
\setlength{\tabcolsep}{38pt}
\begin{center}
\begin{tabular}{lcc}
\hline\hline
\textbf{Parameter} & \textbf{Prior} & \textbf{Value} \\
\hline
\textbf{TOI-4336 A b Parameters} & & \\
Orbital period, $P_b$ (days)\dotfill  & $\mathcal{N}(16.336351, 0.000017)$ & $16.336351 \pm 0.000017$\\
Transit epoch, $T_{0,b}$ (RJD)\dotfill & $\mathcal{N}(59335.57275, 0.00046)$ & $59335.57274 \pm 0.00046$\\
RV semi amplitude, $K_b$ (m/s)\dotfill & $\mathcal{U}(0, 10)$ & $1.85^{+0.18}_{-0.20}$ \\
Eccentricity, $e_b$\dotfill & Fixed & $0.0$ ($<$ 0.45, 3$\sigma$) \\
Argument of periastron, $\omega_b$ (deg)\dotfill & Fixed & $90.0$ \\
\textbf{TOI-4336 A c Parameters} & & \\
Orbital period, $P_c$ (days)\dotfill  & $\mathcal{N}(7.587266, 0.000012)$ & $7.587266 \pm 0.000012$\\
Transit epoch, $T_{0,c}$ (RJD)\dotfill & $\mathcal{N}(59333.2931, 0.0015)$ & $59333.2931 \pm 0.0015$\\
RV semi amplitude, $K_c$ (m/s)\dotfill & $\mathcal{U}(0, 10)$ & $1.114^{+0.093}_{-0.094}$ \\
Eccentricity, $e_c$\dotfill & Fixed & $0.0$ ($<$ 0.35, 3$\sigma$) \\
Argument of periastron, $\omega_c$ (deg)\dotfill & Fixed & $90.0$ \\
\textbf{Gaussian Process Parameters} & & \\
$A_1$ (m/s) \dotfill & $\mathcal{U}(0, 10)$ & $4.2^{+2.5}_{-1.4}$ \\
$A_2$ (km/s) \dotfill & $\mathcal{U}(0, 10)$ & $1.24^{+0.51}_{-0.32}$ \\
$B_2$ (K) \dotfill & $\mathcal{U}(0, 10)$ & $0.0078^{+0.0036}_{-0.0020}$ \\
$O_{\mathrm{amp},1}$  \dotfill & log$\mathcal{U}(0.001, 10)$ & $1.98^{+1.4}_{-0.77}$ \\
$O_{\mathrm{amp},2}$  \dotfill & log$\mathcal{U}(0.001, 10)$ & $0.33^{+0.13}_{-0.15}$ \\
$P_\mathrm{dec}$ (days) \dotfill & $\mathcal{U}(40, 300)$ & $68^{+26}_{-18}$ \\
$P_\mathrm{rot}$ (days) \dotfill & $\mathcal{U}(10, 100)$& $35.5 \pm 1.1$ \\
\textbf{Instrument Offsets and Jitters} & & \\
$\sigma_{RV}$ (m/s) \dotfill & $\mathcal{U}(0.001,62)^\star$ & $0.16^{+0.11}_{-0.10}$ \\
$\mu_{RV}$ (m/s) \dotfill & $\mathcal{U}(7938,27948)^\star$  & $17943.42 \pm 2.5$ \\
$\sigma_{\Delta T_{3500K}}$ (K)  \dotfill & $\mathcal{U}(0.001,20)^\star$  & $0.410^{+0.084}_{-0.064}$ \\
$\mu_{\Delta T_{3500K}}$ (K) \dotfill & $\mathcal{U}(-10000,10000)^\star$  & $0.34^{+0.52}_{-0.46}$ \\
$\sigma_{FWHM}$ (km/s) \dotfill & $\mathcal{U}(0.00,0.07)^\star$  & $0.00264^{+0.00046}_{-0.00036}$ \\
$\mu_{FWHM}$ (km/s) \dotfill & $\mathcal{U}(-10000,10000)^\star$  & $4.9741^{+0.0031}_{-0.0030}$ \\
\hline
\end{tabular}
\begin{tablenotes}
\item
\textbf{Notes.} $\mathcal{N}(\mu, \sigma^{2})$ indicates a normal distribution with mean $\mu$ and variance $\sigma^{2}$, $\mathcal{U}(a, b)$ a uniform distribution between $a$ and $b$, log$\mathcal{U}(a, b)$ a log-uniform distribution between $a$ and $b$. $^\star$ Automatically determined by \texttt{PyOrbit}.
\end{tablenotes}
\label{tab:posteriors_RV_TOI4336}
\end{center}
\end{table*}

\begin{figure}[h]
\centering
    \includegraphics[width=0.82\textwidth]{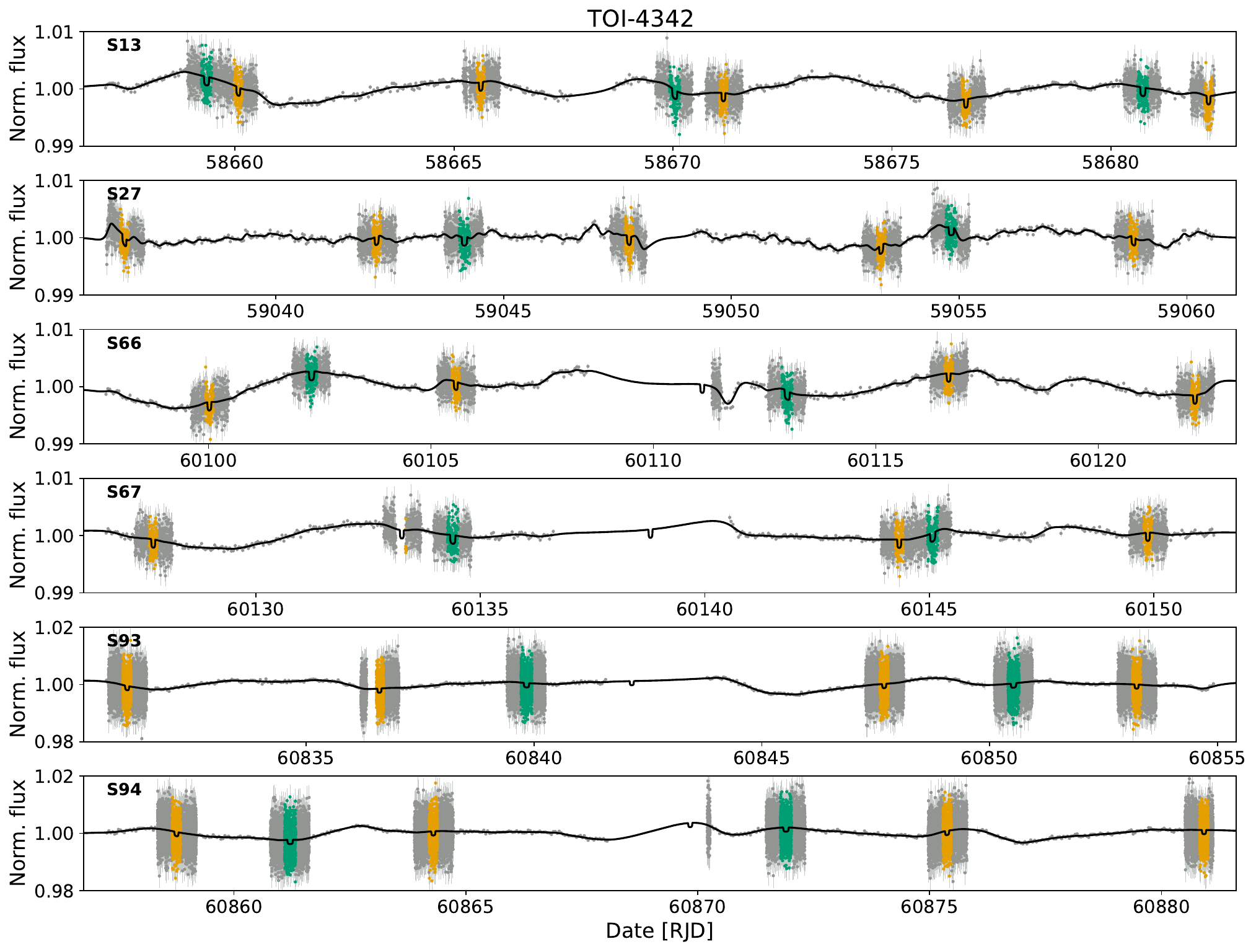}
     \caption{Normalized TESS PDCSAP flux light curves of TOI-4342 from the six different sectors are shown in gray, with the best-fit model in black (see Sect.~\ref{subsubsect:photometry_TOI4342} for details on the modeling). Transits of planets b and c are highlighted in orange and green, respectively.}
     \label{fig:TESS_full_model_4342}
\end{figure}
\begin{figure}[h]
\centering
    \includegraphics[width=\textwidth]{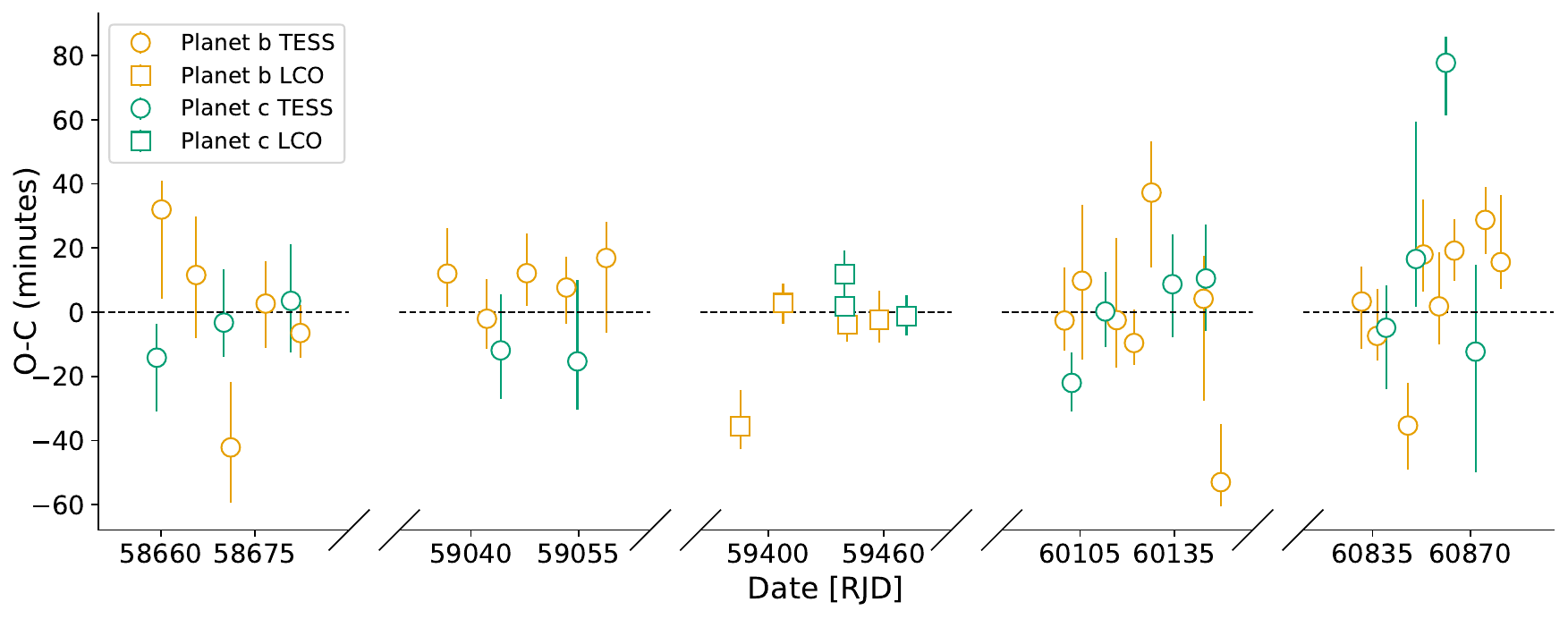}
     \caption{Observed minus calculated transit times for all observed TESS and LCO transits of TOI-4342~b and c.}
     \label{fig:TTVs_4342}
\end{figure}

\begin{figure}[h]
\centering
    \includegraphics[width=0.95\textwidth]{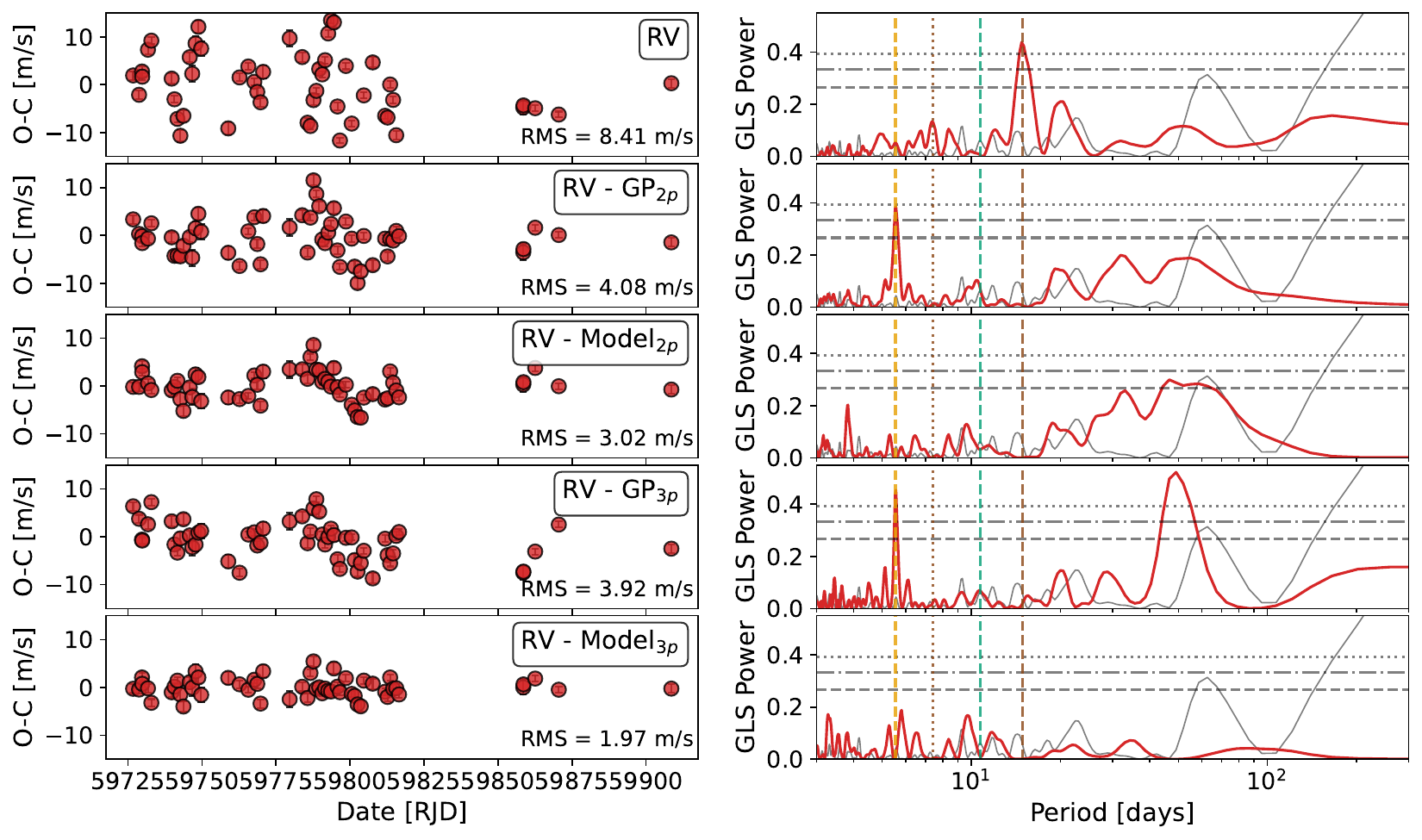}
     \caption{Time series of the RVs (left column) of TOI-4342 and the corresponding periodograms in red (right column) for different residual data sets. From top to bottom: observed RVs; RVs after subtraction of the GP model (2p); RVs after subtraction of the full 2p + GP model; RVs after subtraction of the GP model (3p); and RVs after subtraction of the full 3p + GP model. The vertical brown lines indicate the stellar rotation period at $\sim$14 days (dashed line) and its first harmonic at $\sim$7 days (dotted line). The orange and green vertical dashed lines mark the expected orbital periods of planets b and c, respectively. The 10\%, 1\%, and 0.1\% False Alarm Probability (FAP) levels are shown as gray horizontal dashed, dash-dotted, and dotted lines, respectively.}
     \label{fig:TOI4342_periodograms_residuals}
\end{figure}

\begin{table*}[h]
\tiny
\caption{Median values and 68$\%$ confidence intervals of the posterior distributions of the photometric and RV fit of TOI-4342.}
\centering
\renewcommand{\arraystretch}{1.1}
\setlength{\tabcolsep}{38pt}
\begin{center}
\begin{tabular}{lcc}
\hline\hline
\textbf{Parameter} & \textbf{Prior} & \textbf{Value} \\
\hline
\textbf{TOI-4342 b Parameters} & & \\
Orbital period, $P_b$ (days)\dotfill  & $\mathcal{U}(5.40, 5.70)$ & $5.5382592 \pm 0.0000034$\\
Transit epoch, $T_{0,b}$ (RJD)\dotfill & $\mathcal{U}(58653, 58657)$ & $58654.53479^{+0.00084}_{-0.00093}$\\
Scaled planetary radius, $R_{p,b}$/$R_{*}$ \dotfill & $\mathcal{U}(0, 0.5)$ & $0.03571^{+0.00073}_{-0.00075}$ \\ 
Impact parameter, $b_b$ \dotfill & $\mathcal{U}(0, 1)$ & $0.382^{+0.062}_{-0.063}$ \\ 
RV semi amplitude, $K_b$ (m/s)\dotfill & $\mathcal{U}(0, 10)$ & $3.78^{+0.65}_{-0.67}$ \\
Eccentricity, $e_b$\dotfill & Fixed & $0.0$ \\
Argument of periastron, $\omega_b$ (deg)\dotfill & Fixed & $90.0$ \\
\textbf{TOI-4342 c Parameters} & & \\
Orbital period, $P_c$ (days)\dotfill  & $\mathcal{N}(10.60, 10.80)$ & $10.688662 \pm 0.000015$\\
Transit epoch, $T_{0,c}$ (RJD)\dotfill & $\mathcal{N}(58656, 58665)$ & $58659.3486 \pm 0.0017$\\
Scaled planetary radius, $R_{p,c}$/$R_{*}$ \dotfill & $\mathcal{U}(0, 0.5)$ & $0.03602^{+0.00092}_{-0.00093}$  \\ 
Impact parameter, $b_c$ \dotfill & $\mathcal{U}(0, 1)$ & $0.320^{+0.12}_{-0.13}$ \\ 
RV semi amplitude, $K_c$ (m/s)\dotfill & $\mathcal{U}(0, 10)$ & $1.97^{+0.57}_{-0.56}$ \\
Eccentricity, $e_c$\dotfill & Fixed & $0.0$ \\
Argument of periastron, $\omega_c$ (deg)\dotfill & Fixed & $90.0$ \\
\textbf{Candidate d Parameters} & & \\
Orbital period, $P_d$ (days)\dotfill  & $\mathcal{U}(2, 100)$ & $47.5 \pm 1.3$\\
Transit epoch, $T_{0,d}$ (RJD)\dotfill & $\mathcal{U}(58750, 58850)$ & $58795 \pm 27$\\
RV semi amplitude, $K_d$ (m/s)\dotfill & $\mathcal{U}(0, 20)$ & $4.49^{+0.73}_{-0.75}$ \\
Eccentricity, $e_d$\dotfill & Fixed & $0.0$ \\
Argument of periastron, $\omega_d$ (deg)\dotfill & Fixed & $90.0$ \\
\textbf{Stellar parameters} & & \\
Stellar density, $\rho_{*}$ ($\rho_{\odot}$)  \dotfill & $\mathcal{N}(2.75,0.26)$ & $2.86^{+0.20}_{-0.23}$ \\
$c_{1,TESS}$  \dotfill & $\mathcal{N}(0.30,0.10)$ & $0.280 \pm 0.081$ \\
$c_{2,TESS}$   \dotfill & $\mathcal{N}(0.34,0.10)$ & $0.344 \pm 0.092$ \\
$c_{1,LCO}$   \dotfill & $\mathcal{N}(0.33,0.10)$  & $0.317 \pm 0.087$ \\
$c_{2,LCO}$   \dotfill & $\mathcal{N}(0.31,0.10)$ & $0.297 \pm 0.095$ \\
\textbf{Gaussian Process Parameters} & & \\
$A_1$ (m/s) \dotfill & $\mathcal{U}(0, 100)$ & $8.6^{+2.4}_{-1.7}$ \\
$A_2$ (m/s) \dotfill & $\mathcal{U}(0, 100)$ & $5.4^{+1.4}_{-1.0}$ \\
$A_3$ \dotfill & $\mathcal{U}(0, 10)$ & $0.49^{+0.17}_{-0.11}$ \\
$B_3$ (K) \dotfill & $\mathcal{U}(0, 10)$ & $6.4^{+2.2}_{-1.9}$ \\
$O_{\mathrm{amp},1}$  \dotfill & log$\mathcal{U}(0.01, 10)$ & $0.21^{+0.14}_{-0.17}$ \\
$O_{\mathrm{amp},2}$  \dotfill & log$\mathcal{U}(0.01, 10)$ & $0.048^{+0.096}_{-0.032}$ \\
$O_{\mathrm{amp},3}$  \dotfill & log$\mathcal{U}(0.01, 10)$ & $3.96 \pm 0.70$ \\
$P_\mathrm{dec}$ (days) \dotfill & $\mathcal{U}(20, 200)$ & $79^{+14}_{-13}$ \\
$P_\mathrm{rot}$ (days) \dotfill & $\mathcal{U}(10, 100)$& $14.691^{+0.080}_{-0.070}$ \\
\textbf{Instrument Offsets and Jitters} & & \\
$\sigma_{RV}$ (m/s) \dotfill & $\mathcal{U}(0.001,176)^\star$ & $2.22^{+0.47}_{-0.41}$ \\
$\mu_{RV}$ (m/s) \dotfill & $\mathcal{U}(-13950,6091)^\star$  & $-3922.3 \pm 3.2$ \\
$\sigma_{BIS}$ (m/s) \dotfill & $\mathcal{U}(0.02,850)^\star$  & $1.38^{+1.0}_{-0.9}$ \\
$\mu_{BIS}$ (m/s) \dotfill & $\mathcal{U}(-10000,10000)^\star$  & $39.7^{+1.7}_{-1.8}$ \\
$\sigma_{TESS}$  \dotfill & $\mathcal{U}(0.00,0.04)^\star$  & $0.000378^{+0.000029}_{-0.000027}$ \\
$\sigma_{\Delta T_{4000K}}$ (K) \dotfill & $\mathcal{U}(0.001,30)^\star$  & $0.474^{+0.077}_{-0.070}$ \\
$\mu_{\Delta T_{4000K}}$ (K) \dotfill & $\mathcal{U}(-10000,10000)^\star$  & $1.3^{+4.6}_{-4.4}$ \\
\hline
\end{tabular}
\begin{tablenotes}
\item
\textbf{Notes.} $\mathcal{N}(\mu, \sigma^{2})$ indicates a normal distribution with mean $\mu$ and variance $\sigma^{2}$, $\mathcal{U}(a, b)$ a uniform distribution between $a$ and $b$, log$\mathcal{U}(a, b)$ a log-uniform distribution between $a$ and $b$. $^\star$ Automatically determined by \texttt{PyOrbit}.
\end{tablenotes}
\label{tab:posteriors_RV_TOI4342}
\end{center}
\end{table*}

\clearpage
\twocolumn

\section{Discussion}
\subsection{TESS coverage of possible transits of TOI-4342~d}\label{subsect:tess_planet_d}

The RV modeling of the TOI-4342 system revealed a possible planetary candidate with an orbital period of approximately 47.5 days (see Sect.~\ref{subsubsect:RVs_TOI4342}). In this appendix, we evaluate whether the transits of this planet could have been observed in the available TESS sectors. Due to the limited time span of our RV data, the time of conjunction of its orbit remains poorly constrained, preventing us from phase-folding the light curves at a fixed ephemeris to search for a possible transit.

We first estimate the expected impact parameter under the assumption of coplanarity with planet b ($i_b = 88.83^\circ$) or planet c ($i_c = 89.61^\circ$). Between these two limits, the candidate d would have an impact parameter ranging from $0.53 \pm 0.34$ to $1.59 \pm 0.30$, allowing for a potentially transiting configuration. With a minimum mass of 17.8 $M_\oplus$, we estimate a radius of $\sim$4.2 $R_\oplus$ following the mass–radius relation of \citet{Parc2024}. The corresponding transit depth would be about 4200 ppm, implying that even a single transit would have been detectable, yet no such event is observed in the six available TESS sectors.

\begin{figure}[h]
\centering
    \includegraphics[width=0.48\textwidth]{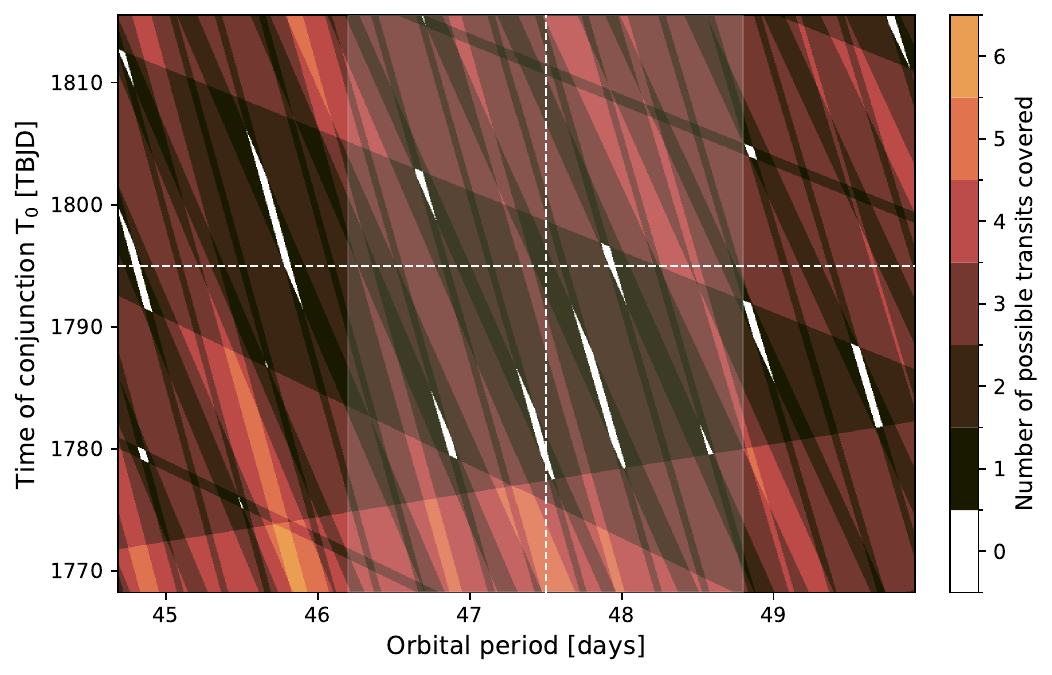}
     \caption{TESS transit coverage map for the candidate planet TOI-4342 d.
    Each pixel corresponds to a combination of orbital period ($P$) and time of conjunction ($T_0$). The color scale indicates the number of possible transits that fall within the TESS observation windows. The dashed white lines mark the median values of $P_d$ and $T_{0,d}$ derived from the RV analysis, while the shaded white bands show the corresponding 1$\sigma$ uncertainties of $P_d$. The regions in white colors correspond to parameter combinations for which not covered by TESS observations.}
     \label{fig:transits_cov_TOI4342_d}
\end{figure}

Given the relatively long orbital period of the candidate, the transit may have occurred in-between the observed TESS sectors. To quantify this, we constructed a grid in ($P$, $T_0$) space, using the minimum and maximum values of the period posterior derived in Sect.~\ref{subsubsect:RVs_TOI4342}, and assessed the TESS coverage for each combination. We find that 98.8\% of the possible transits are covered by the available TESS data, leaving only a 1.2\% chance that the planet could have transited outside the TESS observing window. This result is illustrated in Fig.~\ref{fig:transits_cov_TOI4342_d}, where the white regions correspond to combinations of ($P$, $T_0$) not covered by TESS observations. Fortunately, these remaining low-probability cases will be tested in the near future, as TESS is scheduled to observe TOI-4342 for three months in 2026 (Sectors 101–103, from 1 March 2026 to 17 May 2026).

\subsection{Detailed interior modeling}
\label{app:interior}

We analyzed the potential composition of the planets in more detail using an inference model based on \citet{Dorn2015} with updates in De Wringer et al. (in review). The inference uses a physical forward model based on \citet{dorn_generalized_2017} with updates described in \citet{dorn2021hidden, luo_majority_2024}.
The underlying forward model consists of three layers: an iron core, a silicate mantle, and a H$_2$-He-H$_2$O or a pure H$_2$O steam atmosphere.
For the solid phase of the iron core, we used the equation of state (EOS) of hexagonal close packed iron \citep{hakim_new_2018,miozzi_new_2020}. 
For the liquid iron phase, we used the EOS from \citet{luo_majority_2024}. 
The silicate mantle is composed of three major species MgO, SiO$_2$, and FeO. 
We modeled the solid phase of the mantle using the thermodynamical model \textsc{Perple\_X} \citep{connolly_geodynamic_2009} for pressures below $\approx 125\,$GPa, while for higher pressures we defined the stable minerals a priori and used their respective EOS from various sources \citep{hemley_constraints_1992,fischer_equation_2011,faik_equation_2018,musella_physical_2019}.
The liquid mantle was modeled as a mixture of Mg$_2$SiO$_4$, SiO$_2$ and FeO, as there is no data for the density of liquid MgO in the required pressure-temperature regime \citep{melosh_hydrocode_2007,faik_equation_2018,ichikawa_ab_2020,stewart_shock_2020}. 
In all cases, the EOS of the different components were mixed using the additive volume law for ideal gases. 
Both the iron core and the silicate mantle were modeled as adiabatic. 

For the three sub-Neptunes, we consider a H$_2$-He-H$_2$O atmosphere, while for the super-Earth case (TOI-4336~A~c), we consider the possibility of water but not the presence of a primordial gas envelope. 

For sub-Neptunes, we consider the H$_2$-He-H$_2$O atmosphere layer for which we use the analytic description by \citet{guillot_radiative_2010} and \citet{2014_Jin_planetarypopulation}. It consists of an irradiated layer on top of a nonirradiated layer in radiative-convective equilibrium. 
The water mass fraction is given by $Z_\mathrm{env}$, and the hydrogen-helium ratio is set to solar. 
The two components of the atmosphere, H$_2$/He and H$_2$O, were again mixed following the additive volume law. 
We used the EOS by \citet{1995_Saumon_EOS} for H$_2$/He and the ANOES EOS \citep{1990_thompson_aneos} for H$_2$O. The transit radius of the planet is defined at the radius where the chord optical depth is $\tau_\text{ch}=0.56$.

The transit radius of a planet with a steam envelope is assumed to be at a pressure of $P_{\rm Transit}=1$ mbar. This is a simplification as the transit radius depends on temperature, however, the effect on the planet of interest is small. The thermal profile is assumed to be fully adiabatic, except for pressures less than the pressure at the tropopause (here fixed at 0.1 bar) where we keep an isothermal profile that equals the equilibrium temperature.

Water can be added to the mantle and core melts, depending on its solubility and partitioning behavior, for which we follow \citep{dorn_hidden_2021, luo_majority_2024}. The addition of water reduces the density of the mantle and core melts, for which we follow \citet{bajgain_structure_2015} and decrease the melt density per wt\% water by $0.036$ g cm$^{-3}$. For small water mass fractions, this reduction is nearly independent of pressure and temperature. The addition of water in core melts lowers the density as described in \cite{luo_majority_2024}. The effect of dissolved water on melting temperature is accounted for.
Beyond dissolved water in the deep interior, water can be in solid, supercritical and steam phase, for which we employ the EOS compilation in \citet{haldemann_aqua_2020}. 

For the inference, we used a Polynomial Chaos Kriging surrogate modeling \citep{DeWringer} to approximate the global behavior of the full physical forward model and replace it in the MCMC framework \citep{schobi2015polynomial,marelli2014uqlab}. For this inference, the surrogate model provides high quality fits with mostly R-squared values (coefficient of determination) of $>$0.999 for the data of planetary mass, radius, and bulk Fe/Si ratio. Also, the surrogate root mean square errors are well below observational uncertainties with values between 0.0003 -- 0.03 for planetary mass and between 0.002--0.01 for planetary radius, and 0.06 for the bulk Fe/Si ratio. Those errors of the model uncertainty are accounted for in the likelihood function.

\begin{table}[h]
\centering
\small
\caption{Inference priors for sub-Neptunes (top) and super-Earths (bottom).}
\label{tab:priors_inference}
\renewcommand{\arraystretch}{1.}
\begin{tabular}{lc}
\hline\hline
\textbf{Parameter} & \textbf{Prior} \\
\hline \\[-6pt]
$M_\mathrm{atm} \; (M_\oplus)$ & $\ln \mathcal{N}(-2,1)$ \\
$M_\mathrm{core+mantle} \; (M_\oplus)$ & $\mathcal{N}(M_p, \sigma_{M_p}^2)$ \\
$Z_\mathrm{env}$ & $\mathcal{U}(0.02, 1.0)$ \\
\hline \\[-6pt]
$M_\mathrm{water} \; (M_\oplus)$ & $\mathcal{U}(0.0, 0.05)\times M_p^*$ \\
$M_\mathrm{mantle} \; (M_\oplus)$ & $\mathcal{U}(0.1, 0.99)\times M_p$ \\
$M_\mathrm{core} \; (M_\oplus)$ & $\mathcal{U}(0.1, 0.99)\times M_p$ \\
\bottomrule
\end{tabular}
\tablefoot{$^*$Used only if water is considered.}
\end{table}

\begin{figure}[h]
\centering
    \includegraphics[width=0.4\textwidth]{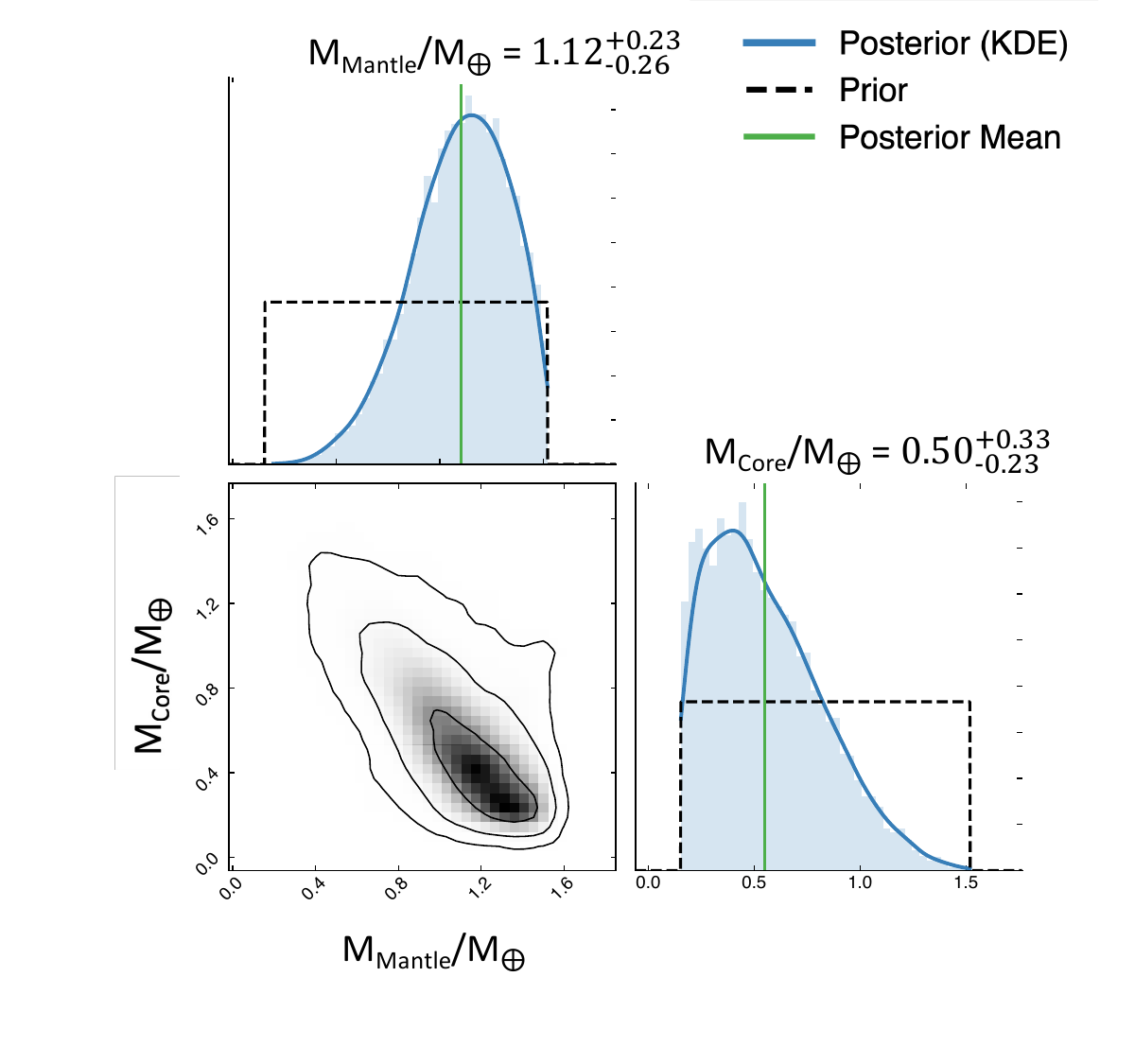}
    \includegraphics[width=0.39\textwidth]{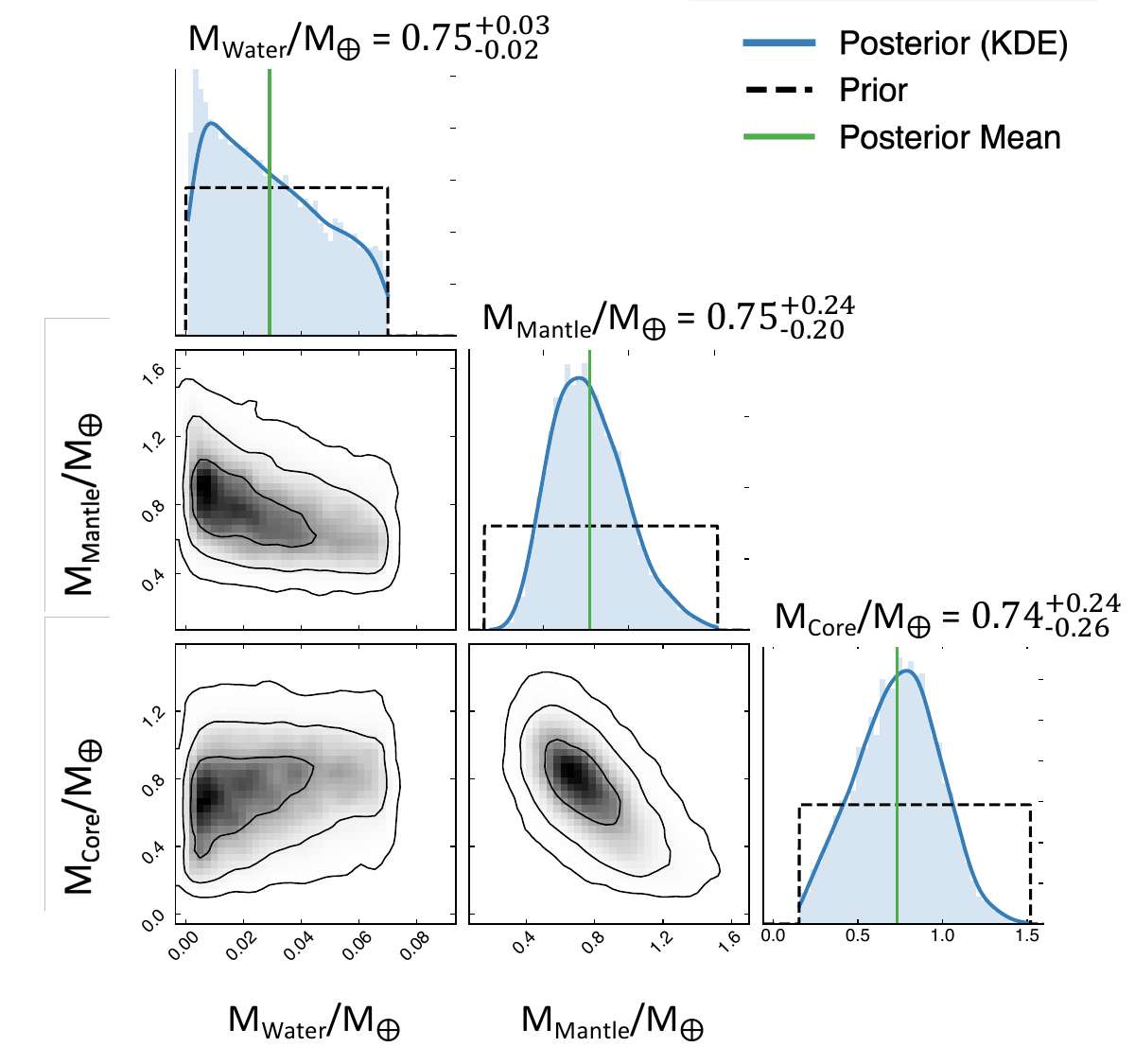}
     \caption{1-D and 2-D marginalized posteriors for the super-Earth TOI-4336~A~c: for a pure rocky world (top) and a water-containing world, while allowing for dissolved water in the deep interior (bottom).}
     \label{fig:int_post_4336bc}
\end{figure}

\begin{figure}[h]
\centering
    \includegraphics[width=0.43\textwidth]{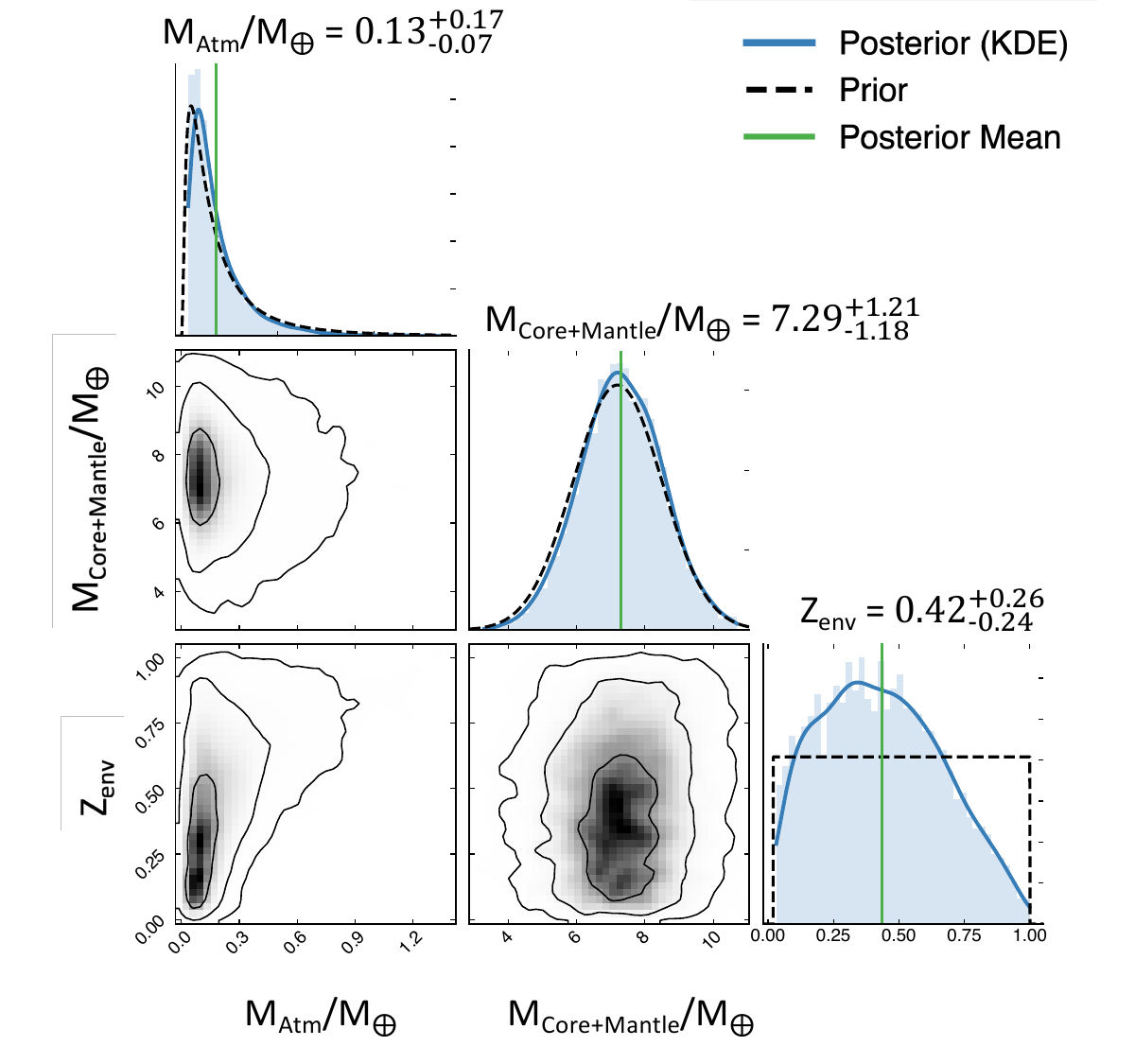}
    \includegraphics[width=0.43\textwidth]{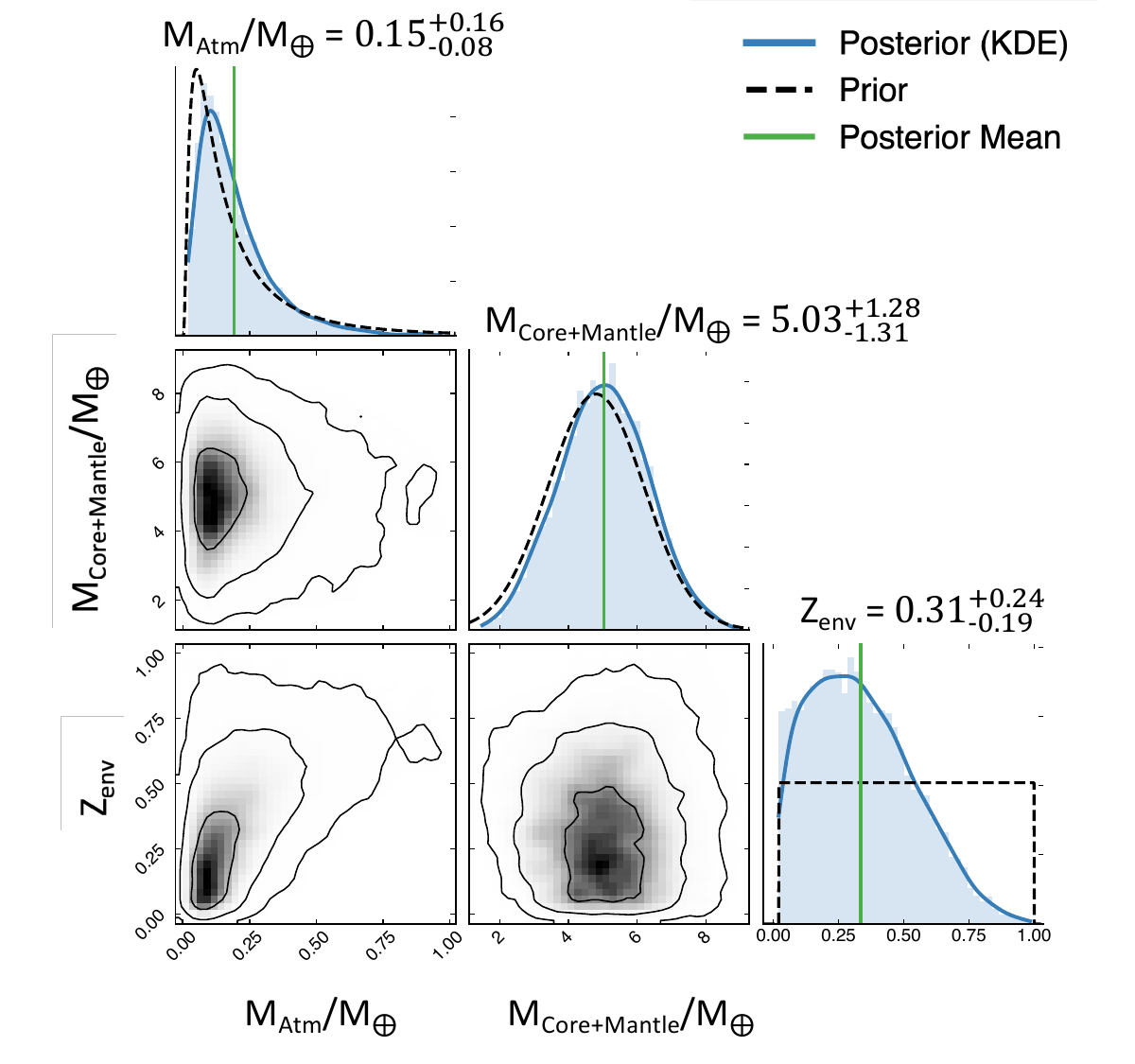}
    \includegraphics[width=0.43\textwidth]{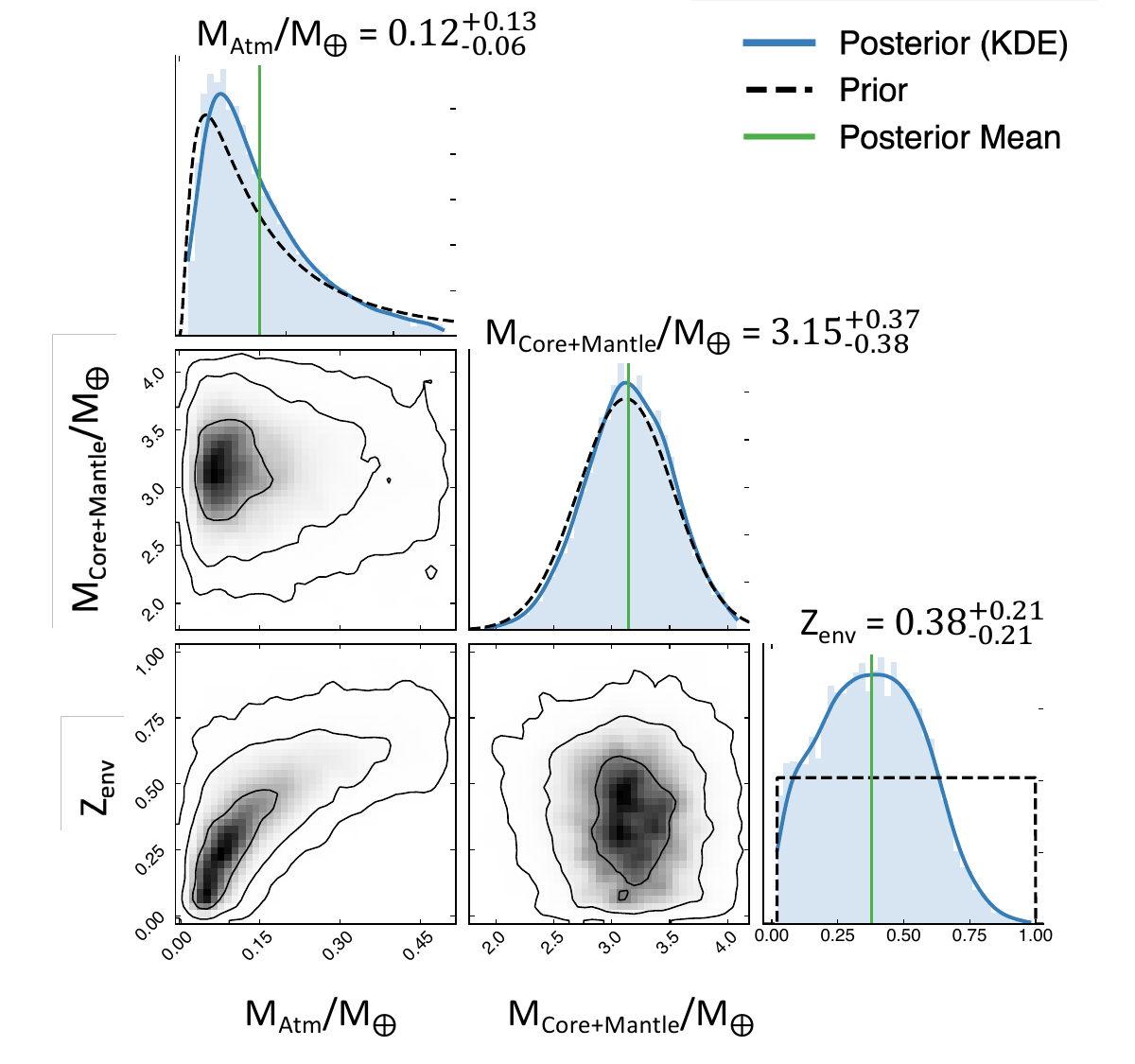}\\
     \caption{1-D and 2-D marginalized posteriors for the sub-Neptunes TOI-4342~b (top), TOI-4342~c (middle), and TOI-4336~A~b (bottom).}
     \label{fig:int_post_4342bc}
\end{figure}

\clearpage

\end{appendix}

\end{document}